\documentclass[12pt]{article}
\title{}
\author{}
\date{}
\usepackage[utf8]{inputenc}
\usepackage[letterpaper,top=1in,right=1in,left=1in,bottom=1in]{geometry}
\usepackage{amsmath}
\usepackage{amsfonts}
\usepackage{amssymb}
\usepackage{graphics}
\usepackage[demo]{graphicx}
\usepackage{placeins}
\usepackage[capposition=top]{floatrow}
\usepackage{subcaption}
\usepackage{lscape}
\usepackage{pdflscape}
\usepackage{setspace}
\usepackage{standalone}
\usepackage{fancyhdr}
\usepackage{framed}
\usepackage{color}
\usepackage{tikz}
\usepackage{pgfplots}
\usepackage{longtable}
\usepackage{booktabs,caption}
\usepackage[font=singlespacing]{caption}
\usepackage[draft]{todonotes}
\usepackage[flushleft]{threeparttable}
\usepackage{bbm}
\usepackage{multirow}
\usepackage{lineno}
\usepackage{titlesec}
\usepackage{hyperref}
\usepackage{comment}
\usepackage[draft]{todonotes}
\usepackage[justification=centering]{caption}
\pgfplotsset{compat=1.15}

\usepackage{longtable}
\usepackage{array}
\usepackage{authblk}
\usepackage{pdflscape}
\usepackage{colortbl}
\usepackage{ragged2e}
\usepackage{xfp}
\usepackage{lipsum}

\usepackage{stackengine}
\usepackage{array}
\newcolumntype{L}[1]{>{\raggedright\arraybackslash}p{#1}}
\setstackEOL{\#}
\setstackgap{L}{12pt}

\usepackage[page,toc,titletoc,title]{appendix}
\usepackage{blindtext}
\usepackage{pdfpages}

\hypersetup{
	colorlinks=true, %set true if you want colored links
	linktoc=all,     %set to all if you want both sections and subsections linked
	linkcolor=black,  %choose some color if you want links to stand out
	citecolor=black,
    urlcolor=blue
}

\usepackage[style=apa,natbib]{biblatex}
\usepackage[all]{nowidow} % Tries to remove widows
\usepackage[protrusion=true,expansion=true]{microtype} % Improves typography, load after fontpackage is selected
\usepackage{longtable}
\usepackage{orcidlink}

\usepackage{sectsty}
\allsectionsfont{\normalsize}

\hypersetup{ 	
pdfsubject = {},
pdftitle = {IL-HTE Economics},
pdfauthor = {Gilbert, Joshua B.}
}

\usepackage{setspace}

\title{Estimating Heterogeneous Treatment Effects with Item-Level Outcome Data: Insights from Item Response Theory}

\begin{document}

\begin{titlepage}
\author[1]{Joshua B. Gilbert\,\orcidlink{0000-0003-3496-2710}}
\author[1]{Zachary Himmelsbach\, \orcidlink{0000-0002-5444-0648}}
\author[2]{James Soland\, \orcidlink{0000-0001-8895-2871}}
\author[3]{Mridul Joshi\, \orcidlink{0009-0006-4690-1981}}
\author[3]{Benjamin W. Domingue\,\orcidlink{0000-0002-3894-9049}}
\affil[1]{Harvard University Graduate School of Education}
\affil[2]{University of Virginia School of Education and Human Development}
\affil[3]{Stanford University Graduate School of Education}

\maketitle

\begin{abstract}

\noindent Analyses of heterogeneous treatment effects (HTE) are common in applied causal inference research. However, when outcomes are latent variables assessed via psychometric instruments such as educational tests, standard methods ignore the potential HTE that may exist among the individual items of the outcome measure. Failing to account for ``item-level'' HTE (IL-HTE) can lead to both underestimated standard errors and identification challenges in the estimation of treatment-by-covariate interaction effects. We demonstrate how Item Response Theory (IRT) models that estimate a treatment effect for each assessment item can both address these challenges and provide new insights into HTE generally. This study articulates the theoretical rationale for the IL-HTE model and demonstrates its practical value using 75 datasets from 48 randomized controlled trials containing 5.8 million item responses in economics, education, and health research. Our results show that the IL-HTE model reveals item-level variation masked by single-number scores, provides more meaningful standard errors in many settings, allows for estimates of the generalizability of causal effects to untested items, resolves identification problems in the estimation of interaction effects, and provides estimates of standardized treatment effect sizes corrected for attenuation due to measurement error. \\
\end{abstract}



\noindent Corresponding author: \href{mailto:joshua_gilbert@g.harvard.edu}{joshua\_gilbert@g.harvard.edu} \\

\newpage

\footnotesize

\noindent \textbf{Funding}: This research was partially supported by the U.S. Department of Education, Institute for Education Sciences, Grant R305D220046. The opinions expressed are those of the authors and do not represent the views of the Institute or the U.S. Department of Education (JG). This research was partially supported by the Jacobs Foundation (BD). \\

\noindent \textbf{Data and code availability}: Our data, code, results, and supplemental analyses are available at the following URL: \url{https://doi.org/10.7910/DVN/C4TJCA}. We have also uploaded cleaned versions of our datasets to the Item Response Warehouse \citep[\url{https://redivis.com/datasets/as2e-cv7jb41fd}]{domingue2023item}, available with the prefix \texttt{gilbert\_meta}. The original datasets are publicly available. Our replication materials contain URLs to both the original articles and replication packages for each study, as well as the raw and cleaned datasets for convenience. \\

\noindent \textbf{Acknowledgments}: The authors wish to thank Alex Bolves, Mauricio Romero, Peter Halpin, Andrew Ho, Jeffrey Smith, two anonymous reviewers, and seminar participants at Harvard University, Stanford University, University of Oslo, University of California Berkeley, Pontifical Catholic University of Chile, Modern Modeling Methods, the International Meeting of the Psychometric Society, and the Applied Statistics Workshop at the Institute for Quantitative Social Science for their helpful comments on this paper. \\

\noindent \textbf{Author Contributions}:

\noindent Conceptualization: JG, BD

\noindent Methodology: JG

\noindent  Software: JG

\noindent Formal analysis: JG

\noindent  Writing---original draft preparation: JG

\noindent  Writing---review and editing: JG, ZH, JS, MJ, BD

\end{titlepage}

\doublespacing

\section*{INTRODUCTION}

Analytic methods to assess heterogeneous treatment effects (HTE) play a critical role in program and policy evaluation. They reveal for whom and under what conditions an intervention works \citep{schochet2014understanding, kent2018personalized, xie2012estimating, wager2018estimation, imai2013estimating, bryan2021behavioural}. One common approach to HTE analyses is to interact treatment indicators with pretreatment covariates to estimate whether treatment efficacy differs by person or site characteristics. However, when the outcome is a latent variable \citep{bollen2002latent} assessed using a psychometric instrument such as an educational test, psychological survey, or patient self-report of disease symptoms (e.g., \cite{koretz2008measuring, jacob2016measurement, hieronymus2019influence, jessen2018improving, bergenfeld2023measurement, cochran2019latent}), the traditional treatment-by-covariate interaction approach ignores an alternative heterogeneity: variation among the individual items of the outcome measure in their sensitivity to treatment. In this study, we demonstrate that explicitly modeling this potential ``item-level'' HTE (IL-HTE) can both resolve causal identification challenges and provide estimates of statistical precision that account for the sampling of items from a broader domain. In addition, modeling IL-HTE can provide new insights into causal analyses in general. By ignoring IL-HTE, researchers conducting causal analyses of outcome data derived from psychometric instruments can potentially reach incomplete conclusions about the efficacy and policy implications of specific interventions.

To address these challenges, emerging scholarship in education, psychometrics, and epidemiology has proposed item response theory (IRT, \cite{hambleton2013item}) modeling approaches that allow for unique treatment effects on each assessment item \citep{sales2021effect, ahmed2023heterogeneity, gilbert2023jebs, gilbert2024ssri}. Models for IL-HTE can offer new insights into causal inference, identification, and generalizability for both methodologists and applied researchers, although they have not yet been widely applied in policy analysis or program evaluation contexts. The purpose of this study is to articulate the rationale for the IL-HTE approach and demonstrate its wide applicability, illustrated through 75 item-level datasets from 48 randomized controlled trials (RCTs) containing 5.8 million item responses in education, economics, and health research. Using these data, we explore five insights that the IL-HTE model provides for HTE analyses and illustrate the implications of each insight both theoretically and empirically.

We briefly summarize five insights the IL-HTE model provides for the applied researcher analyzing randomized evaluations with psychometric outcome measures. First, the IL-HTE model provides an interpretable measure of variation in item-specific or subscale effects potentially masked by a conventional analysis of a single-number outcome such as a sum score. Second, the IL-HTE model provides the flexibility to account for uncertainty arising from the random sampling of items from a population or superpopulation of items, thus providing inference for the items that \textit{could} have been included in the outcome measure, not just those that \textit{were} included in the outcome measure, if desired. Third, the IL-HTE model allows for out-of-sample generalizations of treatment effects on untested items through prediction intervals covering a range of treatment effects on items drawn from a similar population of items that could have been included in the outcome measure. Fourth, the IL-HTE model can disentangle HTE that depends on item characteristics from HTE that depends on person characteristics, which can become conflated when treatment effects are correlated with item difficulty. Finally, the IL-HTE model (and all latent variable models) provide estimates of standardized effect sizes that are corrected for attenuation due to measurement error. Our approach therefore complements recent calls in the economics literature and social science more broadly for greater consideration of measurement issues in empirical analyses \citep{almaas2023economics, flake2020measurement}.

The remainder of the study is organized as follows. We begin with a conceptual discussion of the potential causes of IL-HTE and review a potential outcomes framework for observed and latent variables. We then review the traditional treatment-by-covariate interaction effect approach to HTE analysis and introduce the IL-HTE model as an extension. We proceed with a description of our data sources, methods, and results. We conclude with a summary and discussion of the affordances and limitations of the IL-HTE model in applied causal inference research.

\subsection*{Potential Causes of Item-Level Heterogeneous Treatment Effects} \label{il_hte_concept}

Several potential causal mechanisms might lead to IL-HTE. First, IL-HTE can arise when an outcome measure is not fully aligned with the intervention that is the object of study \citep{francis2022treatment, wwc2022handbook}. For example, if a math intervention is focused on fractions, but the test covers all fifth-grade math content, only test items related to fractions may show a treatment effect as a result of this under-alignment. This scenario relates to the ``instructional sensitivity'' of the items on a measure \citep{polikoff2010instructional, naumann2014modeling}. From a psychometric perspective, this issue could be conceptualized as having a subscale or testlet model as the true data-generating measurement model \citep{rijmen2010formal}, where the treatment affects only certain subsections of a test, but the researcher estimates treatment effects using only the overall score, thus potentially leading to incomplete conclusions about treatment efficacy. Evidence of this intervention-assessment alignment issue is abundant in studies showing that treatment effect estimates are much larger, on average, when the researcher conducting the study uses a measure designed for that intervention rather than a standardized instrument \citep{wolf2023making, kim2021measures} and has implications for the widely documented phenomenon of treatment effect ``fade out'' over time \citep{bailey2017persistence, bailey2020persistence, abenavoli2019mechanisms}.

Second, if an intervention takes place over the course of several months or years, the developmental appropriateness of the items may shift over time. For example, some test items from earlier grades (i.e., those that cover content relatively less covered in later grades) might paradoxically become \textit{more} difficult as students move on to other material and may therefore create IL-HTE if this trend is more pronounced in the treatment group. Or, in the case of survey items, developmental changes might affect how strong an indicator of the latent construct an item is. For example, a survey item asking if the respondent ``cries easily'' is unlikely to be an equally meaningful indicator of depression when the respondent is 10 years old compared to 70 \citep{curran2008pooling}, and there is evidence that this item is not equally indicative of depression for boys versus girls \citep{steinberg2006using}. From a technical perspective, measurement invariance failures that have nothing to do with the intervention can nonetheless introduce bias into RCTs \citep{soland2021measurement}.\footnote{In short, ``a measure is invariant when members of different populations who have the same standing on the construct being measured receive the same observed score on the test'' \citep[p. 211]{schmitt2008measurement}.}

Third, in education, IL-HTE could be the product of teaching to the test or ``score inflation'' \citep{koretz2008measuring, koretz2005alignment, barlevy2012pay, jacob2003catching}, whereby test scores can increase without a corresponding increase in latent ability. For example, an intervention focused on test-taking strategies such as process of elimination could have the effect of making multiple choice items easier in the treatment group while open response questions would be unaffected by treatment.

Other potential causes of IL-HTE could arise if the outcome of interest is measured with a psychological or social-emotional survey instrument. Specifically, research documents that response shifts can occur in intervention studies when observed changes in respondents' scores reflect something other than true changes in the attribute of interest, such as when an intervention changes the way treated respondents interpret the items \citep{olivera2023intervention, oort2009formal}. For example, some medical studies on quality of life show that patients with serious health conditions can score better than healthy people, or that patients score better after deterioration of health \citep{groenvold1999anxiety}. Such results indicate that patients may adopt other frames of reference than healthy people, or adopt new perspectives when their health changes. A related explanation is that the intervention makes the treated participants better able to identify the responses the researcher wants to hear. Although less studied, differential response style bias, such as socially desirable responding for the treatment group post-intervention \citep{bolt2009addressing, schneider2018extracting, van2004response} or Hawthorne effects \citep{levitt2011there, tiefenbeck2016magnitude, leonard2008patient} could also lead to IL-HTE.

Note that in some cases, these explanations treat sources of heterogeneity as a nuisance resulting from failures of measurement invariance. However, other explanations such as instructional sensitivity allow us to better understand treatment effects by acknowledging that not all items respond equally to the treatment. Under any of the conditions discussed in this section, IL-HTE cannot be identified or understood without first modeling treatment heterogeneity in the item responses.

\subsection*{Potential Outcomes with Observed and Latent Variables} \label{potential_outcomes}

Consider some outcome $\theta_j$ for person $j$. Using the potential outcomes framework \citep{rubin2005causal, imbens2015causal}, we define the individual causal effect of binary treatment $T_j$ on person $j$ as $\tau_j \equiv \theta_j(1) - \theta_j(0)$, where 1 indicates the treatment counterfactual and 0 indicates the control counterfactual. Because only one counterfactual is observed, $\tau_j$ is in principle unobservable and cannot be identified. Therefore, we define the sample average treatment effect (ATE), which is identifiable from the observed data, as $\overline{\tau} = \frac{1}{n}\Sigma_{j = 1}^n (\theta_j(1) - \theta_j(0))$. We can estimate $\overline\tau$ by calculating the difference in means between the treated and control groups when treatment is randomly assigned, because random assignment ensures that the observed treatment status is independent of the potential outcomes.\footnote{Note that we assume a frequentist perspective on causal inference for the purposes of this study. We therefore concentrate on the variance of our estimators over repeated samples. Bayesian inference, which produces statements regarding one's uncertainty about a parameter within a particular sample or population, is an alternative that is growing in popularity in causal inference (see \cite{li2023bayesian} for a review). We restrict ourselves to the repeated sampling view because it remains the most common among applied researchers and our conclusions do not depend on accepting or rejecting any form of Bayesianism.}

In practice, we can estimate the sample ATE with the following linear regression model, where $T_j$ is an indicator variable for the treatment status of person $j$, $\beta_0$ is the mean of the control group, $\beta_1$ is the difference in means between the groups, and $\varepsilon_j$ is the error term \citep{murnane2010methods, imbens2015causal, angrist2009mostly, rosenbaum2017observation}:

\begin{align}
    \theta_j = \beta_0 + \beta_1 T_j + \varepsilon_j. \label{eq:theta}
\end{align}

When $\theta_j$ is observed, the difference in means approach provided by Equation \ref{eq:theta} is standard. However, $\theta_j$ may be an unobserved or latent variable, which requires the use of a proxy observed variable $Y_j$ in its place \citep{bollen2002latent, kmenta1991latent}. Examples of latent variables and their proxies include an observed student math test score as a proxy for unobserved mathematical ability, or a sum score on a depression scale as a proxy for unobserved patient depression, each constructed from a set of items. Typical approaches to constructing $Y_j$ include the Classical Test Theory sum score ($Y_j = \Sigma_{i=1}^{I}X_{ij}$, where $X_{ij}$ is the response to item $i$ by person $j$ and $I$ is the number of items in the measure;  \cite{lord1968statistical}) or Item Response Theory (IRT) approaches that use more complex weighting schemes to calculate the observed score \citep{hambleton2013item}. 

Thus, in many empirical applications, we estimate the causal effect on the proxy outcome $Y_j$:

\begin{align}
    Y_j = \beta_0 + \beta_1 T_j + \varepsilon_j. \label{eq:sum}
\end{align}
\noindent In the case of classical measurement error in  $Y_j$, the error is absorbed into the residual term $\varepsilon_j$. As such, $\beta_1$ from Equation \ref{eq:sum} still provides an unbiased estimator for the ATE, although the model for the proxy outcome is less efficient than a model of the true outcome $\theta_j$ due to the increased residual variance. For this reason, measurement error in the outcome variable is not typically considered a challenge to causal identification. Consequently, measurement issues have received relatively less attention than other threats to causal identification, such as nonrandom attrition or covariate imbalance (e.g., \cite{wwc2022handbook}), although an emerging body of work has begun to explore the implications of measurement issues for causal inference more generally \citep{jpal2023measurement, shear2024measurement, gilbert2024measurement, soland2022survey, ballou2009test}.

Importantly, measurement error in the outcome is relevant for causal inference because it causes attenuation bias when the outcome variable is standardized, a common practice given that test scores and psychological surveys have no natural scale. This attenuation bias occurs because the residual standard deviation $\sigma_\varepsilon$ is inflated due to measurement error. Therefore, when calculating a standardized effect size such as Cohen's $d$ using the formula $\frac{\beta_1}{\sigma_\varepsilon}$, the resulting effect size from the proxy outcome model is driven towards zero relative to the effect size derived from a model of the true outcome. Attenuation bias can be addressed with both classical corrections and latent variable modeling approaches \citep{gilbert2024measurement}, an issue that we return to in our results.\footnote{In simple linear regression with two (unstandardized) variables, measurement error in the predictor serves to attenuate the slope towards 0, whereas measurement error in the outcome does not create bias but reduces precision and power as the measurement error is absorbed into the residual term, though these simple rules of thumb do not always hold in more complex circumstances \citep{kline2023principles, shear2024measurement, deshon1998cautionary, cole2014manifest}.}

In contrast to a two-step approach in which we first construct some proxy $Y_j$ and use the proxy as the outcome in a regression model, consider an alternative analytic approach that estimates the ATE directly from the responses to individual items of the outcome measure without the need to compute a summary score to be used as a proxy outcome in a separate step. For example, if the items are dichotomous, such as correct or incorrect answers on an educational test, we can use the following model to estimate the ATE directly on the latent outcome $\theta_j$:

\begin{align}
\label{eq:unconditional}
    \text{logit}(\Pr(Y_{ij} = 1)) = \eta_{ij} &= \theta_j + b_i \\ 
    \theta_j &= \beta_0 + \beta_1 T_j + \varepsilon_j \\
    b_i &\sim N(0, \sigma^2_b) \\
    \varepsilon_j &\sim N(0, \sigma^2_\theta).
\end{align}
\noindent Here, the log-odds that response to item $i$ by person $j$ equals 1 is a function of latent outcome $\theta_j$ and item location $b_i$. $\theta_j$ is in turn a function of the control group mean $\beta_0$ and the treatment effect $\beta_1$, analogous to Equation \ref{eq:theta}. The equation for $\theta_j$ can be expanded to include additional predictors, such as covariates or treatment-by-covariate interactions. In educational testing contexts, item location $b_i$ is typically interpreted as item easiness, in that items with higher values of $b_i$ are easier to answer correctly. 

Equation \ref{eq:unconditional} is an explanatory IRT model, an approach with origins in the psychometric literature \citep{wilson2004descriptive, wilson2008explanatory}. While IRT and associated models have a rich tradition in psychometrics, education, and psychology \citep{petscher2020past}, applications in these disciplines have been mostly descriptive \citep{gilbert2023estimating}. However, an emerging literature has begun to apply IRT models to causal inference applications in education, economics, and clinical trials, demonstrating the advantages of the IRT approach to causal analyses \citep{ahmed2023heterogeneity, sales2021effect, gilbert2023estimating, gilbert2023jebs, gilbert2024disentangling, gilbert2024ipd, gilbert2024ssri}.\footnote{Though we rely on an IRT formulation, Equation \ref{eq:unconditional} is similar to a structural equation model (SEM) that simultaneously accounts for both measurement and structural components in a single estimation procedure \citep{kline2023principles}. Continuous analogs to the IRT formulation explored here include the traditional factor analytic model and linear SEM \citep{aigner1984latent, bentler1983simultaneous, muthen2002beyond, skrondal2007latent, lewbel1998semiparametric, kline2023principles}. Note that treatment impacts on a latent variable can also be taken into account through a multigroup IRT model. That is, one can fit a model that constrains item parameters to be equal across control and treatment groups but allows control and treatment groups to have unique means and variances for the latent outcome (in contrast, simply regressing the latent variable on a treatment indicator only shifts the latent mean). In addition to helping produce unbiased treatment effect estimates under certain conditions \citep{soland2022survey}, such models can be adapted to match more complex, quasi-experimental scenarios such as difference-in-differences designs \citep{soland2023item}.}

Whether we use a single-number score $Y_j$ as a proxy for a latent outcome of interest in a two-step procedure, or a latent variable model such as Equation \ref{eq:unconditional} that models treatment effects on the item responses directly in a single step, the approaches described so far ignore the HTE that may exist among participants or among items of the assessment. In other words, all models considered above assume that treatment effects are constant across people and the items of the measure. We can relax this assumption by allowing treatment effects to vary using models that allow for HTE.

\subsection*{Modeling Heterogeneous Treatment Effects (HTE)}  \label{trad_hte}

\subsubsection*{\textit{Person-Level HTE}}

We can extend Equation \ref{eq:unconditional} to allow for HTE by some person-level covariate $X_j$ as follows:

\begin{align}
    \theta_j = \beta_0 + \beta_1 T_j + \beta_2 X_j + \beta_3 T_j \times X_j + \varepsilon_j.
\end{align}

\noindent In this model, $\beta_1$ is the conditional ATE (CATE) when $X_j=0$ and $\beta_3$ is the HTE parameter. When $\beta_3=0$, the treatment effect is constant across the range of $X_j$; when $\beta_3 \neq 0$, the CATE depends on the value of $X_j$. $X_j$ could include, for example, variables such as age, gender, baseline ability, or other person characteristics. Treatment-by-covariate interaction models serve as the workhorse of HTE analysis in many disciplines \citep{tian2014simple, donegan2012assessing, donegan2015exploring, lee2005micro, athey2017state, schochet2014understanding, gabler2009dealing, gewandter2019demonstrating}, though they are but one approach among many to estimate HTE. Alternatives include quantile regression, mediation, instrumental variables, subgroup analysis, generalizability analysis, and post stratification \citep{schochet2014understanding, tipton2014generalizable}, as well as machine learning approaches \citep{kunzel2019metalearners, gong2021heterogeneous, wang2024effect, athey2019machine, chernozhukov2018sorted, athey2017econometrics, bonhomme2022discretizing}. In this study, we maintain focus on interaction effects because they are the most commonly applied and most widely understood approach to HTE analysis. Furthermore, no matter how flexibly we estimate the CATE as a function of person or site covariates, all such methods ignore the possibility of IL-HTE.

\subsubsection*{\textit{Item-Level HTE}} \label{il_hte}

As an alternative to the person-centered approach, we can similarly allow for item-level HTE by including a unique treatment effect on each item in the model. Specifically, we extend Equation \ref{eq:unconditional} to include an interaction between treatment and item through the $\zeta_i T_j$ term:
\begin{align}
\label{eq:il_hte_mod}
    \text{logit}(\Pr(Y_{ij} = 1)) = \eta_{ij} &= \theta_j + b_{i} + \zeta_i T_j \\ 
    \theta_j &= \beta_0 + \beta_1 T_j + \varepsilon_j \\
    \begin{bmatrix}
        b_i \\
        \zeta_{i}
    \end{bmatrix}
     &\sim N\left(\begin{bmatrix}
         0 \\ 0
     \end{bmatrix},\begin{bmatrix}
         \sigma^2_b & \rho\sigma_b\sigma_\zeta \\
         \rho\sigma_b\sigma_\zeta & \sigma^2_\zeta
     \end{bmatrix}\right) \\
    \varepsilon_j &\sim N(0, \sigma^2_\theta).
\end{align}

\noindent Here, $\zeta_i$ represents the residual treatment effect on item $i$---i.e., an effect that is above and beyond that of the ATE $\beta_1$. That is, if $\zeta_i>0$, item $i$ shows a larger treatment effect than the average item in the outcome measure and the total treatment effect on item $i$ is equal to $\beta_1 + \zeta_i$. Variation in item-specific treatment effects is parameterized by $\sigma_\zeta$, which reflects the standard deviation (SD) of item-specific treatment effects around the average $\beta_1$. The $\zeta_i$ are equivalent to uniform differential item functioning (DIF) effects caused by the treatment, in that they reflect additional treatment effects on item $i$ after the ATE on $\theta_j$ has been accounted for \citep{gilbert2023jebs, montoya2020mimic, gilbert2024ssri}. As discussed previously, while DIF, and by implication IL-HTE, is traditionally viewed as a nuisance or as evidence of a potential failure of measurement invariance (or both) \citep{olivera2023intervention, shear2024measurement}, we argue that the IL-HTE model can be informative because the unique content of some items may be truly more sensitive to a given treatment and thus revealing of a fine-grained profile of treatment efficacy, rather than indicative of a defective measure \citep{gilbert2024ipd, sukin2010item, gilbert2023jebs}. 

The correlation $\rho$ represents the association between item location $b_i$ and item-specific treatment effect $\zeta_i$. Thus, $\rho>0$ suggests that easier items are more responsive to treatment compared to more difficult items, and vice versa for $\rho < 0$. While $\rho$ may be of interest in itself \citep{gilbert2023jebs}, it is perhaps most relevant in that it can cause identification problems in the estimation of treatment-by-covariate interaction effects when omitted from the model \citep{gilbert2024disentangling}, an issue that we return to in our results.

While IRT analyses conventionally treat items as fixed, our approach to IL-HTE analysis uses random effects for items \citep{holland1990sampling, de2008random}. Econometric analyses of clustered data often prefer fixed to random effects specifications due to concerns about the assumption that the random effects are uncorrelated with the predictors in the model, though when this assumption is met, random effects models are more efficient than their fixed effects counterparts \citep{rabe2022multilevel, bell2019fixed, wooldridge2010econometric, antonakis2021ignoring}. The random effects assumption is not relevant in our proposed modeling approach because in most causal inference applications, the covariates of interest are at the item or person level. Confounding due to violations of the random effects assumption only affects variables that vary across each person-item combination, for example, response time \citep{lu2023mixture}.\footnote{In our application, $T_j$ is constant within person $j$; $T_{ij}$ would suggest that a person received treatment on some items but not others. While such a treatment is conceivable, such as a computerized test that offers additional prompts on some subset of the items, it is unlikely to be the norm in most applied settings. However, with the advent of digital interventions, such designs may become more common and are therefore a promising area for future research \citep{kim2021measures}.} Furthermore, the random effects assumption can be relaxed through a Mundlak specification that includes cluster means as covariates \citep{mundlak1978pooling, antonakis2021ignoring, rabe2022multilevel, curran2011disaggregation, wooldridge2010econometric, wooldridge2019correlated}, and extensions of the Mundlak approach apply to the models discussed here with only minor adjustments \citep{guo2023disaggregating, baltagi2023two}. Random effects models also typically assume a normal distribution on the random effects terms. However, model results tend to be robust to violations of this assumption \citep{knief2021violating, schielzeth2020robustness, bell2019fixed}.

The random effects approach also provides several benefits for our intended application. While in principle we could fit similar models with item fixed effects by interacting each item indicator with the treatment indicator $T_j$, or separate linear or logistic regression models for each item, we argue that the random effects approach is more appropriate in general for addressing IL-HTE. \textcite[pp. 894-895]{gilbert2023jebs} provide five arguments for the use of item random effects, summarized here. First, with item fixed effects, variables representing item characteristics (e.g., subscale, modality, content area) cannot be included in the model as they are collinear with item indicators. Second, each fixed item intercept and item-by-treatment interaction adds a parameter to the model, whereas the random effects approach only requires two additional parameters, the SD $\sigma_\zeta$ and the correlation $\rho$. Thus, the random effects specification is more parsimonious and preserves degrees of freedom when the assumptions underlying the random effects model are met, leading to greater statistical efficiency. Third, the random effects approach provides a direct estimate of the variability of IL-HTE in the data in $\sigma_\zeta$, a value that would be biased upward by measurement error if estimated in a fixed effects approach. Fourth, as a shrinkage estimator, empirical Bayes estimates of item random effects minimize total error and are more stable than their fixed effects counterparts, unless sample sizes are very large, and the empirical Bayes estimation effectively controls for inflated false positive rates due to multiple comparisons \citep{sales2021effect}. Finally, the conceptualization of items as representative of a broader pool of potential items that could have been selected for an assessment---either literally, as in large-scale standardized tests that draw from large item banks for each test administration, or hypothetically, as in a researcher-developed vocabulary assessment that could have plausibly selected different words---matches the random effects approach \citep{jacob2016measurement, brennan1992generalizability, de2008random, koretz2008measuring}.\footnote{Critically, separate or pooled linear probability models for each item would misidentify constant effects on the latent outcome as IL-HTE because a constant treatment effect size on the latent outcome yields varying effect sizes in percentage points that depend on the control group accuracy rate on each item \citep{domingue2022ubiquitous}. In our view, this would represent a statistical artifact of the non-linearity inherent in models of binary outcomes rather than true IL-HTE.}

\section*{METHODS} \label{methods}

\subsection*{Data}

We use 75 publicly available datasets from 48 RCTs containing 5.8 million item responses in economics, education, and related fields. Inclusion criteria for our analysis are as follows: the dataset must include (1) at least 100 subjects, (2) item-level outcome data, (3) a baseline measure prior to intervention (either a lagged outcome or a similar metric to the final outcome)\footnote{While measurement error in pretest variables can create bias in the coefficients of other variables in the regression model, this is more of an issue for estimating average treatment effects in observational rather than experimental studies because any bias is expected to be equal between the treated and control groups; see \textcite{lockwood2014correcting} for a discussion. The degree to which pretest measurement error affects estimates of person-dependent HTE is more complex and will in general attenuate observed HTE. The effects of pretest measurement error on IL-HTE is less studied and is in our view a promising area of future research. One approach to addressing pretest measurement error is to parameterize the pretest as a latent variable when pretest items are available. However, this option is not possible in the \texttt{lme4} software we use in this study, but new advances such as the \texttt{galamm} package \citep{sorensen2024multilevel} may make such models more easily estimable in the future (they are not currently estimable with \texttt{galamm}; Ø. Sørensen, personal communication, December 8, 2024).}, (4) sufficient empirical information in the article or replication materials to verify the scoring of the items to determine whether any items needed to be reverse-coded, and (5) a license that allows sharing of the original data so that researchers can replicate our results and conduct derivative analyses of these datasets. We identified our data sources by first examining existing studies of IL-HTE in education and epidemiology and then expanded our search to replication materials from studies published in the American Economic Review and Economic Journal using a keyword search that included the terms ``randomized controlled trial,'' ``[cluster] randomized trial,'' ``randomized evaluation,'' ``impact evaluation,'' and ``field experiment.'' We focus on these two economic journals because they have strong editorial policies to encourage the publication of data alongside the articles. We also searched Google Dataset Search, Inter-university Consortium for Political and Social Research (ICPSR), the Jameel Poverty Action Lab (JPAL), Innovations for Poverty Action (IPA), and International Initiative for Impact Evaluation (3ie) Dataverse websites for replication materials. We concluded our search in August 2024. Our examples are not intended to be exhaustive but rather illustrative of the advantages of IL-HTE analysis across a broad range of empirical and disciplinary settings.

We make the following simplifications to each dataset in our cleaning to allow the highest degree of comparability between studies and for clarity of exposition. Our replication materials contain clearly documented cleaning code for each study that shows where and how each of the following rules is applied.

\begin{enumerate}
    \item If multiple treatments are administered, we combine groups to represent any treatment (1) compared to any control (0).
    \item We use only the initial treatment assignment as our treatment indicator variable.
    \item We consider randomization as if it were conducted at the individual level, although we note that our models could easily be expanded to include randomization blocks, cluster-corrected standard errors, or multilevel models that include random intercepts for higher-level units such as schools.
    \item If multiple time points are available, we use only the first post-treatment follow-up.
    \item If studies contain multiple outcomes, such as academic test scores and a social-emotional survey, we consider each outcome in separate models. This separation is critical to ensure that we do not conflate IL-HTE in a single construct with HTE across constructs.
    \item For studies reporting only summary measures of baseline or pretest variables, we standardize the pretest variable to mean 0 variance 1 in the sample. If item-level data are available for the pretest measure, we create pretest scores using a Rasch or one-parameter logistic (1PL) IRT model to match our outcome model, which is also a Rasch model.
    \item For the 19 datasets in which the item responses include more than two response categories (e.g., Likert scales ranging from strongly disagree to strongly agree), we convert polytomous responses to dichotomous responses (e.g., strongly agree / agree = 1, disagree / strongly disagree = 0). While it is possible to extend the IL-HTE model to polytomous applications  \citep{gilbert2024ssri}, we dichotomize to make the models directly comparable across all studies. Furthermore, we fit analogous ordinal models for the relevant datasets in our supplement and we see that the substantive findings are essentially unchanged.
\end{enumerate}

\noindent Our analyses therefore depart from many of the original analyses which included design features such as multiple treatments, randomization blocks, additional time points, or other covariates, to maintain focus on the consequences and interpretation of IL-HTE. Our analysis is therefore intended to be illustrative of the affordances of the IL-HTE model across a range of contexts rather than providing a direct replication of the analytic approach used in each original study. We return to extensions of the modeling approach demonstrated here in our discussion.

Table \ref{tab:study_desc} provides descriptive information for each dataset. Our replication materials contain URLs for the original publications and the replication materials for each study, as well as additional description of each intervention.

\begin{footnotesize}
\begin{longtable}{>{\raggedright\arraybackslash}p{6cm}  
                  >{\raggedright\arraybackslash}p{2cm} 
                  >{\raggedright\arraybackslash}p{4.25cm} 
                  >{\raggedright\arraybackslash}p{0.75cm} 
                  >{\raggedright\arraybackslash}p{0.5cm} 
                  >{\raggedright\arraybackslash}p{0.5cm}}

\caption{Descriptive statistics for the datasets in our analysis} \label{tab:study_desc}
\\
\hline
\multicolumn{1}{l}{Dataset} & \multicolumn{1}{l}{Location} & \multicolumn{1}{l}{Outcome} & \multicolumn{1}{l}{$N$}& \multicolumn{1}{l}{$I$} & \multicolumn{1}{l}{$\alpha$}\\ 
\hline
\endfirsthead

\hline
\multicolumn{1}{l}{Dataset} & \multicolumn{1}{l}{Location} & \multicolumn{1}{l}{Outcome} & \multicolumn{1}{l}{$N$}& \multicolumn{1}{l}{$I$} & \multicolumn{1}{l}{$\alpha$}\\ 
\hline
\endhead

\hline
\multicolumn{6}{r}{{Continued on next page}} \\
\endfoot

\hline
\multicolumn{6}{p{\textwidth}}{%
\setstretch{1.0}% Set single spacing for the notes
Notes: N = number of subjects, I = number of items, $\alpha$ = the internal consistency of the test. The letters after the study names do not indicate different publications, but different outcome measures from the same publication. Measures marked with an asterisk indicate that lower values are preferred. The original patient data from \cite{gilbert2024ssri} are private, but the authors provide a simulated dataset of item responses derived from their empirical results that we use in our analysis. \cite{kim2021improving}b and 2021c represent different vocabulary tests administered to Grade 1 and Grade 2 students, respectively. \cite{banerji2017impact}a and b are outcomes for mothers; c and d are outcomes for children. \cite{banerjee2017proof} report several outcomes and samples within the same study.%
}
\endlastfoot

% Your data entries here

1: \cite{gilbert2023jebs}& USA & Reading comprehension & 7797 &  30 & 0.93 \\ 
2:  \cite{kim2023longitudinal}& USA & Reading comprehension & 2174 &  20 & 0.84 \\ 
3: \cite{gilbert2024ssri}a & Europe & Depression (HDRS-17)*& 5314 &  17 & 0.86 \\ 
4: \cite{blattman2017reducing}& Liberia & Crime and violence*& 916 &  20 & 0.94 \\ 
5: \cite{woods2021cluster}& UK & Health literacy & 2486 &   7 & 0.72\\ 
6:  \cite{bruhn2016impact}& Brazil & Financial literacy & 15395 &  10 & 0.69 \\ 
7: \cite{kim2024time}a& USA & Vocabulary & 1352 &  36 & 0.94 \\ 
8: \cite{kim2024time}b& USA & Reading comprehension & 1303 &  29 & 0.89\\ 
9: \cite{hidrobo2016effect}& Ecuador & Domestic violence* & 1284 &  19 & 0.97 \\ 
10:   \cite{kim2021improving}a& USA & Reading self concept & 4834 &  20 & 0.93\\ 
11:   \cite{kim2021improving}b& USA & Vocabulary & 2565 &  24 & 0.86\\ 
12:   \cite{kim2021improving}c& USA & Vocabulary & 2580 &  24 & 0.89 \\ 
13:   \cite{romero2020outsourcing}a& Liberia & Literacy & 3381 &  20 & 0.92 \\ 
14:   \cite{romero2020outsourcing}b& Liberia & Math & 3381 &  44 & 0.98 \\ 
15:   \cite{romero2020outsourcing}c& Liberia & Raven's matrices & 3381 &  10 & 0.75 \\ 
16: \cite{de2024limitations}& India& Math& 3202& 32&0.93\\
17: \cite{duflo2024experimental}a& Ghana& Math& 17344& 21&0.93\\
18: \cite{duflo2024experimental}b& Ghana& English& 17344& 21&0.97\\
19: \cite{duflo2024experimental}c& Ghana& Local language& 17331& 21&0.93\\
20: \cite{jayanthi2021improving} & USA & Math & 186 & 93 & 0.95\\
21: \cite{davenport2023improving} & USA & Math & 3671 & 13 & 0.96 \\
22: \cite{berry2018impact} & Ghana & Savings attitudes & 5290 & 10 & 0.84 \\
23: \cite{bang2023efficacy} & USA & Math & 886 & 38 & 0.95 \\
24: \cite{llaurado2014edal} & Spain & Healthy lifestyle & 495 & 13 & 0.65\\
25: \cite{schreinemachers2020nudging}a & Nepal & Healthy food preferences& 775 & 15 & 0.81\\
26: \cite{schreinemachers2020nudging}b & Nepal & Food knowledge & 775 & 15 & 0.55 \\
27: \cite{banerji2017impact}a & Nepal & Language & 8552 & 18 & 0.94\\
28: \cite{banerji2017impact}b & Nepal & Math & 8552 & 16 & 0.93\\
29: \cite{banerji2017impact}c & India & Language & 14576 & 15 & 0.95\\
30: \cite{banerji2017impact}d & India & Math & 14576 & 10 & 0.95\\
31: \cite{duflo2015wide} & India & Academic achievement & 11893 & 6 & 0.96 \\
32: \cite{maruyama2022strengthening} & El Salvador & Math & 3619 & 20 & 0.87 \\
33: \cite{aladysheva2017impact} & Kyrgyzstan & Social trust & 1242 & 18 & 0.46\\
34: \cite{angelucci2024economic} & India & Depression (PHQ-9)*& 887 & 9 & 0.88 \\
35: \cite{persson2020does}a & Sweden & Democratic values & 1152 & 12 & 0.54 \\
36: \cite{persson2020does}b & Sweden & Political knowledge & 1108 & 7& 0.44\\
37: \cite{carpena2024entertainment}a& India & Health knowledge & 839 & 21 & 0.82  \\
38: \cite{carpena2024entertainment}b& India & Health behavior & 839 & 9 & 0.53  \\
39: \cite{berry2022student}a& Malawi & Cognitive skills & 6196 & 10& 0.80\\ 
40: \cite{berry2022student}b& Malawi & Computation & 6188 & 20& 0.90 \\
41: \cite{mohohlwane2023reading} & South Africa & Oral reading fluency & 3068 & 134 & 0.99 \\
42: \cite{hussam2022psychosocial}a & Bangladesh & Mental health & 726 & 20 & 0.87 \\
43: \cite{hussam2022psychosocial}b & Bangladesh & Cognitive skills & 726 & 25& 0.88\\
44: \cite{lee2022effectiveness} & Hong Kong & Mental health & 219 & 22 & 0.94 \\
45: \cite{bateman2020death} & USA & Burnout/depression* & 173 & 37 & 0.96 \\
46: \cite{glatz2023dynamic}a & Netherlands & Reading & 120 & 42 & 0.93 \\
47: \cite{glatz2023dynamic}b & Netherlands & Math & 123& 44& 0.90\\
48: \cite{cardenas2023parent} & Mexico & Child development & 1150 & 30 & 0.92 \\
49: \cite{daniel2017kusamala} & Malawi & Psychological distress* & 160 & 20 & 0.95 \\
50: \cite{luo2019using}a & China & Parenting beliefs & 449 & 11 & 0.42\\
51: \cite{luo2019using}b & China & Feeding practices & 390 & 8 & 0.38 \\
52: \cite{abaluck2022impact} & Bangladesh & COVID symptoms* & 102873 & 11 & 0.86 \\
53: \cite{saha2023study}a & India & Women's empowerment & 1880 & 14 & 0.40 \\
54: \cite{saha2023study}b & India & Consent & 1880 & 6 & 0.98 \\
55: \cite{wang2023delivering} & Bangladesh & Academic achievement & 1704 & 15 & 0.92 \\
56: \cite{sebele2023improving} & Liberia & Literacy & 2307 & 4 & 0.62 \\
57: \cite{maselko2020effectiveness}a & Pakistan & Depression (PHQ-9)* & 889 & 9 & 0.94 \\
58:  \cite{maselko2020effectiveness}b & Pakistan & Depression (SCID)* & 889 & 13 & 0.98 \\
59:  \cite{maselko2020effectiveness}c & Pakistan & Anxiety (PSS)* & 889 & 10 & 0.90 \\
60: \cite{lyall2020can} & Afghanistan & Government support & 1747 & 9 & 0.58 \\
61: \cite{lyall2020can} & Afghanistan & Violence* & 1747 & 6 & 0.73\\
62: \cite{wilson2024economic} & Kenya & Depression (PHQ-4)* & 709 & 4 & 0.87 \\
63: \cite{hoffmann2018poverty} & India & Indebtedness* & 8987 & 4 & 0.92 \\
64: \cite{zhao2023impacts} & Jordan & Social-Emotional Learning & 4041 & 9 & 0.85 \\
65: \cite{o2023evaluation}a & Australia & Mental Health Literacy & 279 & 14 & 0.90  \\
66: \cite{o2023evaluation}b & Australia & Help Seeking & 272 &  10 & 0.91 \\
67: \cite{o2023evaluation}c & Australia & Mental Health Stigma & 273 & 4 & 0.96  \\
68: \cite{banerjee2017proof}a & India & Hindi & 5974 & 35 & 0.95 \\
69: \cite{banerjee2017proof}b & India & Math & 5966 & 30 & 0.97 \\
70: \cite{banerjee2017proof}c & India & Hindi & 3543 & 24 & 0.96 \\
71: \cite{banerjee2017proof}d & India & Math & 3448 & 20 & 0.99 \\
72: \cite{banerjee2017proof}e & India & Hindi & 2669 & 35 & 0.97 \\
73: \cite{banerjee2017proof}f & India & Math & 2682 & 30 & 0.97 \\
74: \cite{gilbert2024ipd}b & USA & Vocabulary & 1225 & 12 & 0.89 \\
75: \cite{baron2021couples} & USA & Political Engagement & 111 & 10 & 0.83 \\
 \hline
\end{longtable}
\end{footnotesize}

\subsection*{Models} \label{models}

We estimate the following five specifications using mixed effects logistic regression applied to each dataset, presented below in reduced form, where $\eta_{ij}=\text{logit}(\Pr(Y_{ij}=1))$:

\begin{align}
  \label{eq:mod1}  \eta_{ij} =& \beta_0 + \beta_1 T_j + b_i + \varepsilon_j \\
  \label{eq:mod2}   \eta_{ij} =& \beta_0 + \beta_1 T_j + \beta_2 X_j + b_i + \varepsilon_j \\
 \label{eq:mod3}    \eta_{ij} =& \beta_0 + \beta_1 T_j + \beta_2 X_j + \zeta_i T_j + b_i + \varepsilon_j \\
 \label{eq:mod4}    \eta_{ij} =& \beta_0 + \beta_1 T_j + \beta_2 X_j + \beta_3 T_j \times X_j + b_i + \varepsilon_j \\
\label{eq:mod5}     \eta_{ij} =& \beta_0 + \beta_1 T_j + \beta_2 X_j + \beta_3 T_j \times X_j + \zeta_i T_j + b_i + \varepsilon_j.
\end{align}

\noindent Equation \ref{eq:mod1} includes only the treatment indicator, as both a baseline model for comparison and to estimate the residual standard deviation of $\varepsilon_j$, $\widehat\sigma_\theta$, which represents the pooled SD of the latent variable $\theta_j$, which we use to standardize the estimates from the other models. That is, all figures in our tables and graphs are divided by $\widehat\sigma_\theta$ from Equation \ref{eq:mod1} from each dataset so that they can be interpreted in SD units and compared across both models and studies \citep{briggs2008using, gilbert2023jebs}. Equation \ref{eq:mod2} adds the baseline covariate $X_j$. Equation \ref{eq:mod3} adds IL-HTE as a random slope for treatment at the item level. Equation \ref{eq:mod4} adds an interaction between treatment and the baseline covariate, without IL-HTE, to represent the standard person-centered approach to HTE analysis in the IRT framework. Equation \ref{eq:mod5} fits a flexible approach that includes both the treatment by baseline covariate interaction and IL-HTE to allow for both person- and item-dependent HTE \citep{gilbert2024disentangling}. 

The complete hierarchical form of Equation \ref{eq:mod5} is as follows:

\begin{align}
\label{eq:flexible}
    \text{logit}(\Pr(Y_{ij} = 1)) = \eta_{ij} &= \theta_j + b_{i} + \zeta_i T_j \\
    \theta_j &= \beta_0 + \beta_1 T_j + \beta_2 X_j + \beta_3 T_j \times X_j + \varepsilon_j \\ 
    \begin{bmatrix}
        b_i \\
        \zeta_{i}
    \end{bmatrix}
     &\sim N\left(\begin{bmatrix}
         0 \\ 0
     \end{bmatrix},\begin{bmatrix}
         \sigma^2_b & \rho\sigma_b\sigma_\zeta \\
         \rho\sigma_b\sigma_\zeta & \sigma^2_\zeta
     \end{bmatrix}\right) \\
    \varepsilon_j &\sim N(0, \sigma^2_\theta).
\end{align}
\noindent All other models can be interpreted as constrained versions of Equation \ref{eq:flexible}. Compared to Equation \ref{eq:flexible}, Equation \ref{eq:mod4} constrains $\sigma_\zeta$ to 0, Equation \ref{eq:mod3} constrains $\beta_3$ to 0, Equation \ref{eq:mod2} constrains both $\sigma_\zeta$ and $\beta_3$ to 0, and Equation \ref{eq:mod1} constrains $\sigma_\zeta, \beta_3,$ and $\beta_2$ to 0. We provide a directed acyclic graph for Equation \ref{eq:flexible} in Appendix \ref{dag}.\footnote{Equation \ref{eq:flexible} is a generalized linear mixed model (GLMM) with a logistic link function and random effects for items and persons. GLMMs can be fit using various statistical programs including Stata (\texttt{melogit}), SPSS (\texttt{PROC NLMIXED}), R (\texttt{glmer}), and Mplus. We conduct our analyses in R, and we include basic R syntax to fit each model in Appendix \ref{r_code} \citep{gilbert2023tutorial, de2011estimation}.}

\section*{RESULTS} \label{results}

We focus on the broad patterns of results across studies to highlight the interpretation and implications of IL-HTE for empirical analysis. Our supplement shows the full regression output and fit statistics for each model applied to each dataset. We categorize our results in terms of five insights provided by the IL-HTE model, each representing implications for causal inference, identification, or generalizability. Unless otherwise specified, the results and figures below are drawn from Equation \ref{eq:mod3} to maintain focus on IL-HTE considered separately from simultaneous analysis of person-dependent HTE.

\subsection*{The IL-HTE Model Reveals Variation in Item-specific or Subscale Effects Potentially Masked by the ATE} \label{blups}

\begin{figure}
    \centering
    \includegraphics[width=0.95\linewidth]{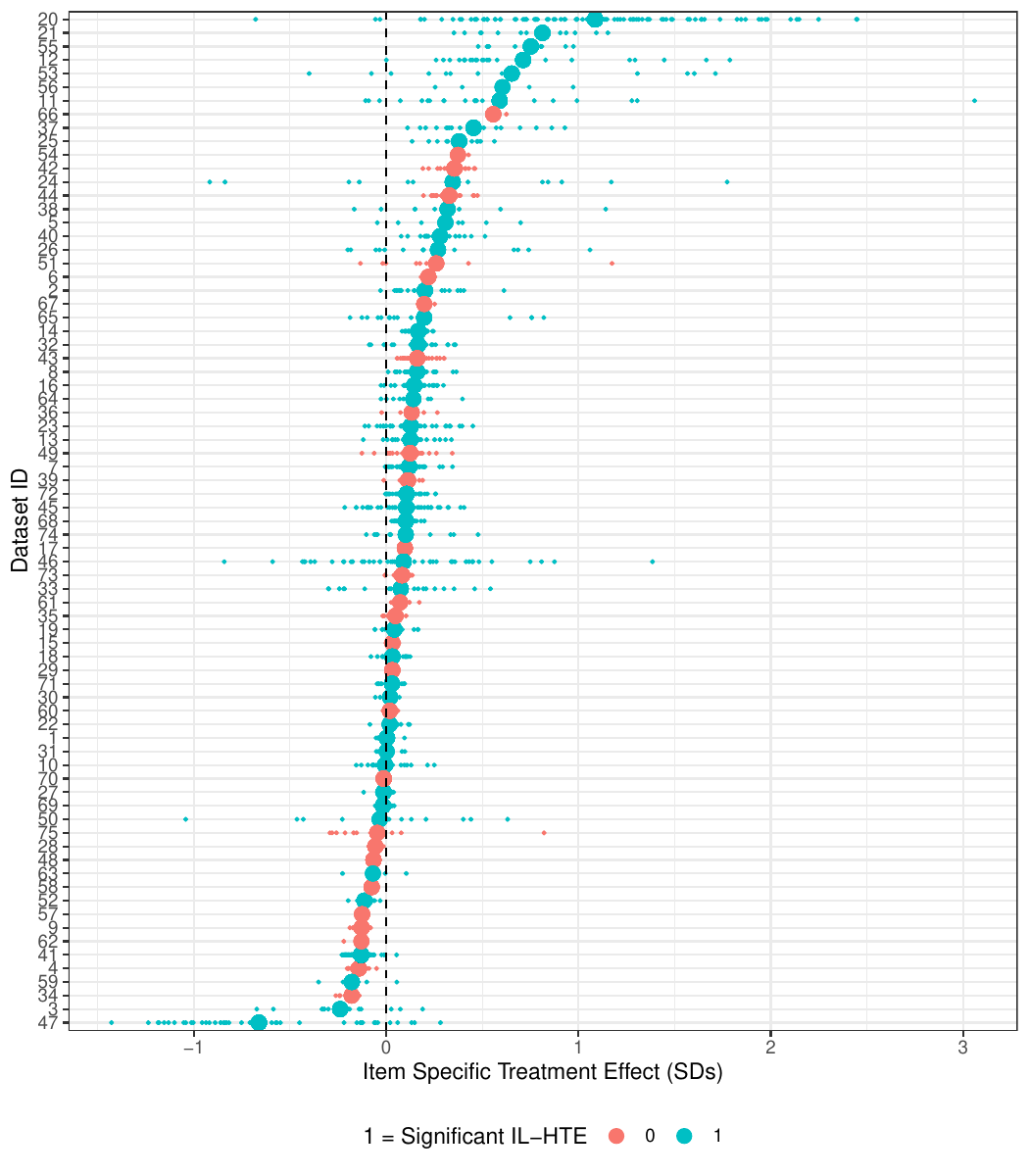}
    \caption{Empirical Bayes estimates of item-specific treatment effects}
            \justify \footnotesize
Notes: The figure shows the distribution of empirical Bayes estimates of item-specific treatment effects by dataset. The small points represent individual item effects and the large points indicate study mean effects. The points are color coded by whether the IL-HTE variance ($\sigma^2_\zeta$) is statistically significant at the 5\% level. The statistical significance of each item-specific treatment effect can also be calculated, and prior work shows that the empirical Bayes method for estimating item-specific effects is robust to multiple comparisons \citep{sales2021effect}. We show graphs of item-specific effects based on separate linear probability models for each item in our supplement.
    \label{fig:item_blups}
\end{figure}

A single-number estimate of the ATE may mask substantively meaningful variation in item-specific or subscale effects. To illustrate the range of item-specific treatment effects in our data, Figure \ref{fig:item_blups} shows the distribution of ATEs ($\beta_1$) and empirical Bayes estimates of item-specific effects ($\beta_1 + \zeta_i$) by study. The points are color-coded by whether the IL-HTE variance ($\sigma^2_\zeta$) is statistically significant at the 5\% level, derived from a likelihood ratio test.\footnote{Because the null hypothesis of 0 variance is on the boundary of the parameter space, the reported $p$-value must be divided by two \citep[p. 5061]{gilbert2023tutorial}.} We see a wide range of distributions of item-specific effects across studies. For example, we see that dataset 20 shows a very large positive ATE and very high IL-HTE. In contrast, dataset 71 shows a near null ATE and essentially no IL-HTE. Dataset 11 shows one outlying item with a large positive effect over 3$\sigma$, suggesting that a content analysis of this item could be informative. This level of fine-grained detail would be obscured in an analysis that did not model IL-HTE and instead considered only a single-number summary score of the outcome measure. 

What explains these disparate patterns of IL-HTE across studies? While the examples in our introduction provide some possibilities in the abstract, explaining IL-HTE in any given context requires substantive knowledge of the nature of the intervention and the outcome measure. For example, \cite{abaluck2022impact} (dataset 52) examine the effect of a masking policy on COVID-19 symptoms and the results show a very narrow range of item-specific effects. In this case, we have a strong theoretical reason to believe that the latent variable model is true (i.e., unobserved COVID-19 causes various symptoms), and it seems plausible that masking would affect all symptoms equally by reducing infections, in contrast to other medical studies where interventions may target both the overall latent outcome and treat specific symptoms that would result in greater IL-HTE. In contrast, consider \cite{kim2021improving}b (dataset 11), who measure vocabulary and show extensive IL-HTE. This result suggests that some vocabulary items are potentially more aligned with the intervention than others, which is again plausible in the context of a literacy intervention that may have emphasized key vocabulary words in intervention lessons. While we can only speculate on what mechanisms underlie the IL-HTE observed for any individual study here, the IL-HTE model provides a framework for quantifying such variation and can serve as an important first step in exploratory or hypothesis-generating analysis and in theory-building more generally \citep{borsboom2021theory}.

In our supplement, we provide an exploratory analysis to shed light on this issue by regressing $\widehat\sigma_\zeta$ on various dataset characteristics. We find that, for example, higher internal consistency ($\alpha$) and the use of standardized outcome measures (vs. researcher-developed measures) are statistically significantly associated with lower IL-HTE, suggesting that IL-HTE may be related to underlying psychometric features of the outcome measure. A more detailed content analysis of each intervention and outcome measure using our datasets could be a fruitful avenue for future research to predict what types of interventions or measures tend to yield greater or lesser IL-HTE (see \cite{gilbert2024mechanisms}).

Along these lines, we view a further promising application of the IL-HTE model to be the estimation of treatment-by-item characteristic interactions or subscale effects. We consider a subscale to be a collection of items from a common content domain within a larger scale that presents as unidimensional (though others use subscale to refer to multidimensional contexts) \citep{stone2009providing, sinharay2018subscores, edwards2024book}. For example, by interacting treatment status with an item-level variable, such as whether an item represents algebra or geometry subscales on a test of overall mathematics proficiency, we can go beyond the unexplained variation of the random slopes model to the systematic variation of an interaction model that allows for more informative tests of specific hypotheses about the potential mechanisms underlying IL-HTE \citep{gilbert2024ipd}. For example, \textcite{gilbert2024ssri} report that selective serotonin reuptake inhibitors (SSRIs) have the strongest impact on the subset of depression items measuring mood rather than more generalized physical symptoms of depression, such as loss of appetite; these findings align with the biochemical mechanisms through which SSRIs affect brain chemistry. While subscale analyses could in principle be carried out with separate regression models for each subscale, such an approach requires \textit{a priori} selection of the subscale, the assumption that within each subscale, the item effects are constant, offers no direct test of the differences in effects by subscale, and does not adjust for the lower reliability of subscales measured with fewer items than the overall scale \citep{gilbert2023tutorial}.  We provide an example formula for an IL-HTE model with subscale effects and some additional discussion in Appendix \ref{subscale} and include R code to fit such a model in Appendix \ref{r_code}.\footnote{While testlet models provide an alternative to the treatment-by-subscale interactions described here, we view the IL-HTE approach as a more general method that is more appropriate in many empirical settings. For example, a testlet approach would likely be a poor model to test for differential treatment effects on, say, multiple choice vs. open response items, or the position of an item in a test, each of which could be included in the IL-HTE model as treatment-by-subscale interactions. Treatment-by-subscale interactions also allow us to calculate pseudo-$R^2$ values by comparing $\sigma^2_\zeta$ from a model without the interactions to one with the interactions to determine how the residual IL-HTE may be reduced. Previous subscale analyses have shown large explanatory power of subscale interactions; e.g., pseudo-$R^2$ values of $90\%$ when allowing differential effects by reading comprehension passage \citep{gilbert2023jebs} and $50\%$ when the mood-level subscale was allowed to vary relative to physical symptoms of depression \citep{gilbert2024ssri}.}

\subsection*{The IL-HTE Model Provides Standard Errors That Account for the Selection of Items in the Construction of the Outcome} \label{SEs}

Standard errors from models that do not allow for IL-HTE ignore the variation caused by the selection of items onto an assessment. That is, models that assume that the treatment effects are constant across the items allow for only one source of uncertainty, namely, sampling variation at the person level. If treatment differentially impacts each item, the variability related to the selection of a specific set of items used to measure outcomes also matters. Had a different set of items been selected, the estimated treatment effect would be different in each realization of the test, holding the people constant \citep{brennan1992generalizability, gleser1965generalizability, gilbert2023jebs, gilbert2024ipd}. As a random slopes model, the IL-HTE approach explicitly accounts for the sampling error of the items selected onto the assessment and provides SEs that account for this added uncertainty \citep{gilbert2023jebs, bell2019fixed, gilbert2024ssri}. We argue that the SEs provided by the IL-HTE model are therefore more meaningful given that in most applications the focus should be on making causal inferences to the population of items that \textit{could} have been on an outcome measure, rather than the specific items included in a single realized outcome administration. This is especially true when estimating causal effects on standardized educational assessments where specific items change over time, such as the SAT, ACT, NAEP, or state standardized tests, or on health outcomes such as hearing loss or depression where different symptoms could have been selected for measurement \citep{gilbert2024ssri, jessen2018improving}.\footnote{If a set of items on an assessment is truly fixed, or the researcher is only interested in inference for the treatment effect on the administered items, then a model with item fixed effects or random intercepts is appropriate. See \textcite[pp. 894-895]{gilbert2023jebs} for a discussion in the IL-HTE context and \textcite{miratrix2021applied} for a discussion of analogous issues in multi-site trials. Thus, the decision to use the IL-HTE model may involve both statistical and substantive considerations that depend on the intervention context and the inferential goals of the analyst.}

When $\sigma_\zeta$ is large, the estimated SE of the ATE $\beta_1$ increases, sometimes substantially, according to the following formula derived from Generalizability Theory \citep{brennan1992generalizability}, where ${V}(\beta_1)_{\text{Rand. Intercepts}}$ is the variance of $\beta_1$ from a model that assumes a constant treatment effect (i.e., the SEs provided by the random intercepts model in Equation \ref{eq:mod2}) and $I$ is the number of items:

\begin{align}
\label{eq:se_inf}
        {\text{SE}}(\beta_1)_{\text{IL-HTE}} = \sqrt{{V}(\beta_1)_{\text{Rand. Intercepts}} + \frac{\sigma^2_\zeta}{I}}. 
\end{align}

\begin{figure}
    \centering
    \includegraphics[width=1\linewidth]{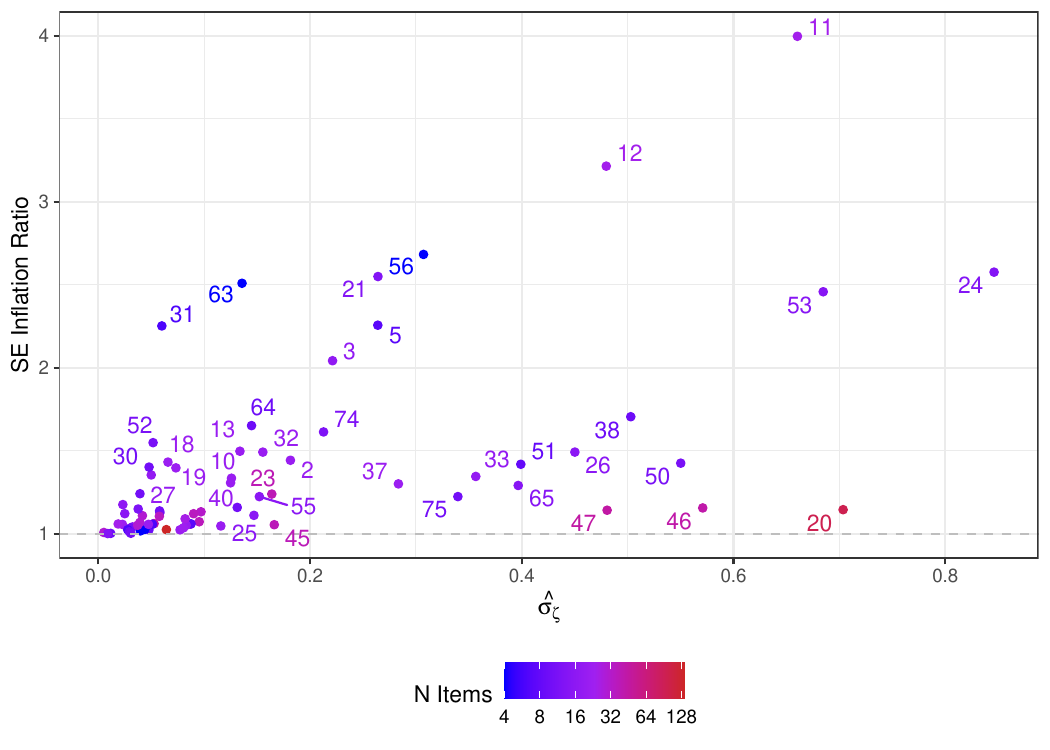}
    \caption{Standard error inflation as a function of IL-HTE}
            \justify \footnotesize
Notes: The vertical axis shows the ratio of SEs from the IL-HTE model to the random intercepts model. That is, a value of 1 indicates that the SEs are equal and a value of 2 indicates that the SE in the IL-HTE model is double that of the random intercepts model. The horizontal axis shows the estimated SD of item-specific treatment effects in each dataset ($\widehat\sigma_\zeta$). The points are labeled by dataset ID and color coded by the number of items.
    \label{fig:se_inflation}
\end{figure}

\noindent The inflation of SEs due to $\sigma_\zeta$ in our data is displayed in Figure \ref{fig:se_inflation}, which plots the ratio of the SE derived from the IL-HTE model (Equation \ref{eq:mod3}) to the SE derived from the constant effects model (Equation \ref{eq:mod2}) against $\widehat\sigma_\zeta$. We observe a strong relationship between this SE ratio and $\widehat\sigma_\zeta$, with the ratio exceeding a factor of 4 in the most extreme case, a large difference that is equivalent to reducing the effective sample size by a factor of 16. While such insights about the inflation of SEs have long been recognized in power analysis and sample size considerations for clustered data generally \citep{killip2004intracluster, abadie2023should, thompson2011simple, wooldridge2003cluster}, consideration of the same principles with regard to assessment items has so far been underappreciated \citep{gilbert2023jebs, ahmed2023heterogeneity}. An important implication of Equation \ref{eq:se_inf} is that when substantial IL-HTE is present, sampling more people will not substantially improve precision, only more items will \citep[Figure S7]{domingue2022intermodel}. 

Furthermore, Equation \ref{eq:se_inf} allows for a novel approach to sensitivity analysis to determine how robust a result is to unobserved IL-HTE \citep{imai2010identification, oster2019unobservable}. That is, for a given statistically significant treatment effect size, SE, and number of items $I$, we can calculate how much IL-HTE would be required to render the result statistically insignificant. Importantly, the sensitivity analysis can be easily calculated without item-level data using the formula presented in Equation \ref{eq:se_inf} by solving for $\sigma_\zeta$ to determine the point at which an estimated treatment effect size would become statistically insignificant (denoted $\Gamma$). Assuming a positive treatment effect, the formula yields (the algebra is presented in Appendix \ref{sens_algebra}):

\begin{align}
     \Gamma &= \sqrt{I\left(\left(\frac{\widehat\beta_1}{1.96}\right)^2 - \widehat{V}(\beta_1)_{\text{Rand. Intercepts}}\right)}. \label{eq:sens}
\end{align}

\noindent As an example, suppose an analysis of a total test score revealed $\widehat{\beta_1} = .3, \widehat{\text{SE}} = .1$. The critical value of $\Gamma$ required to render the effect statistically insignificant is .52 for a 20-item test and .37 for a 10-item test.\footnote{Note that the model for the total score will be on a different scale than a model for the item responses. The present argument holds when the two scales are proportional, which simulation evidence suggests is a plausible assumption \citep{gilbert2023estimating, gilbert2024measurement}. Thus, the Equation \ref{eq:sens} should be viewed as approximate in the total score case.} While large, these figures are consistent with the range of $\widehat\sigma_\zeta$ observed in our data and could thus make many observed effect sizes that ignore this source of variance statistically insignificant, as we will explore in the next section. 

\subsection*{The IL-HTE Model Allows for Out-of-Sample Generalizations}

Related to the SE considerations described earlier, the IL-HTE model also allows us to calculate a prediction interval (PI; \cite{borenstein2023avoiding, patel1989prediction}) that covers a plausible range of expected treatment effects on out-of-sample items. A 95\% PI could have important practical implications in, for example, medical contexts, to identify whether the treatment effects are positive for untested symptoms \citep{gilbert2024ssri}. While a 95\% CI may be far from 0 indicating a statistically significant ATE, a 95\% PI that crosses zero could indicate negative side effects of a treatment. The formula for the PI is as follows \citep[p. 130]{borenstein2009introduction}:
\begin{align}
    \text{PI} = \widehat\beta_1 \pm 1.96\sqrt{\widehat{V}(\beta_1)_\text{IL-HTE} + \widehat\sigma^2_\zeta}.
\end{align}
\noindent Substituting in Equation \ref{eq:se_inf}, we can expand the formula for the PI to

\begin{align}
    \text{PI} = \widehat\beta_1 \pm 1.96\sqrt{\widehat{V}(\beta_1)_{\text{Rand. Intercepts}} + \frac{\widehat\sigma^2_\zeta}{I} + \widehat\sigma^2_\zeta}.
\end{align}
\noindent Note that as both the number of persons and number of items increase, the 95\% PI approaches $\widehat{\beta_1}\pm 1.96 \widehat\sigma_\zeta$.

\begin{figure}
    \centering
    \includegraphics[width=1\linewidth]{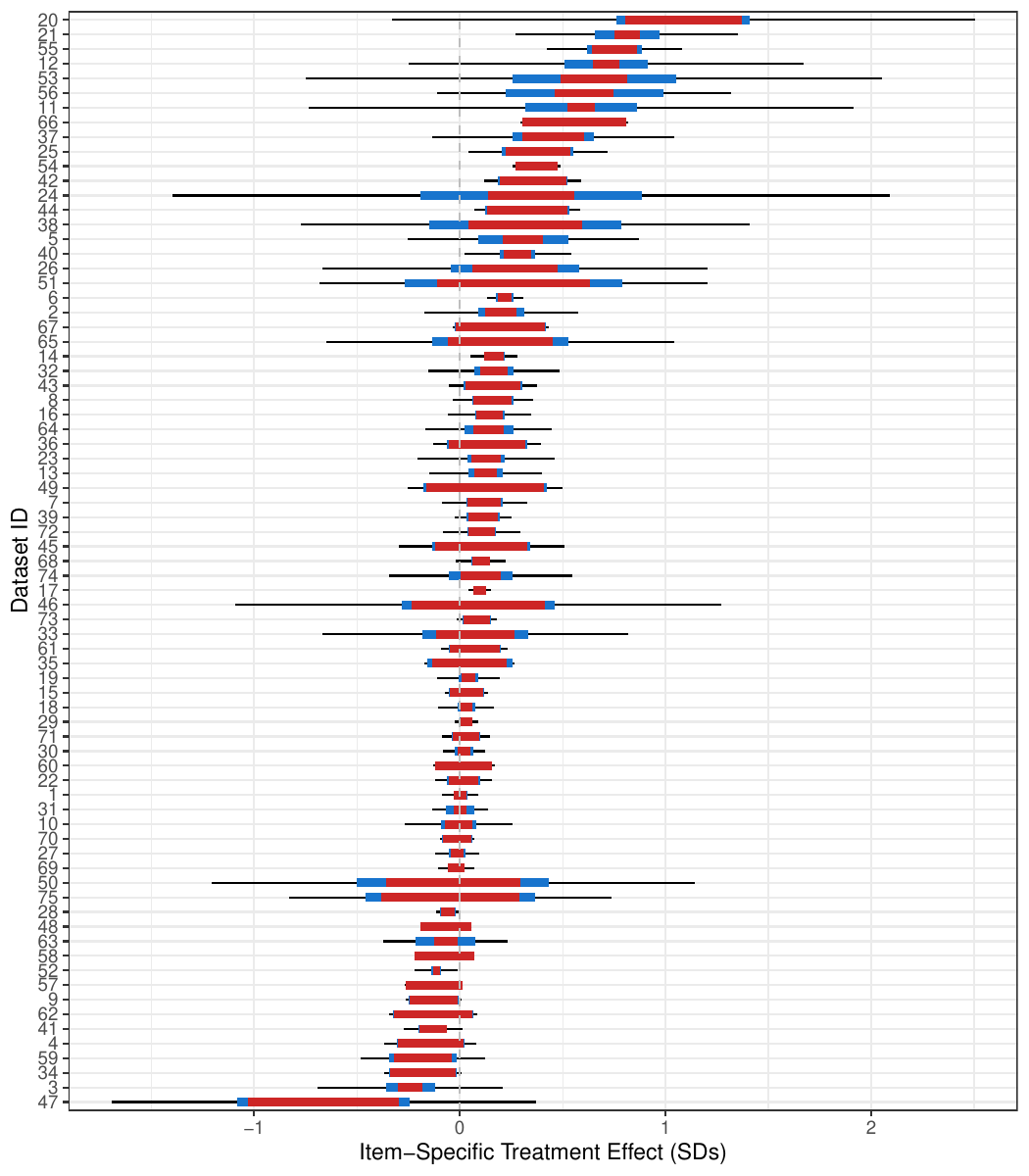}
    \caption{Confidence Intervals (CIs) and Prediction Intervals (PIs) for each study}
            \justify \footnotesize
Notes: The red shows the 95\% CI in the constant effects model. The blue shows the 95\% CI in the IL-HTE model. The black shows the 95\% PI, providing a range of item-specific treatment effects on out of sample items. 
    \label{fig:generalizability}
\end{figure}

Figure \ref{fig:generalizability} shows the 95\% CI for the random intercepts model (red), the IL-HTE model (blue), and the 95\% PI for treatment effects on out-of-sample items (black). We highlight results for three illustrative studies: dataset 6, measuring financial literacy, dataset 11, measuring vocabulary, and dataset 3, measuring depression. Dataset 6 shows a narrow 95\% CI that is barely inflated when IL-HTE is taken into account, and the 95\% PI is also narrow, suggesting that treatment effects on other indicators of financial literacy would likely be similarly affected by the intervention. In contrast, we see in dataset 11 that the narrow 95\% CI from the random intercepts model dramatically increases in the IL-HTE model, and the 95\% PI shows a very large range for out-of-sample vocabulary words, suggesting the intervention effects could vary widely depending on which vocabulary words are tested. As a middle ground, dataset 3 shows a 95\% CI that is moderately inflated by IL-HTE and a 95\% PI that includes positive values, suggesting that some depression symptoms could be worsened by treatment. Whereas Figure \ref{fig:item_blups} shows the observed distribution of item-specific treatment effects in a given study, the 95\% PI provides a range of treatment effects on an untested item drawn from the same population of items---in the present case, other related vocabulary words, other depression symptoms, or other indicators of financial literacy. In sum, the IL-HTE approach provides novel insights on the generalizability of causal effects to untested circumstances.

\subsection*{The IL-HTE Model Resolves Identification Problems in the Estimation of Treatment by Covariate Interaction Effects} \label{identification_problem}

One serious problem associated with ignoring IL-HTE is that it can be impossible to distinguish between person-level and item-level HTE when the analysis focuses on aggregate outcomes. Suppose that there is interest in treatment heterogeneity as a function of some person-level covariate $X_j$. It is possible to estimate a treatment-by-covariate interaction effect (Equation \ref{eq:mod4}) when, in fact, the true model contains no interaction but instead IL-HTE (Equation \ref{eq:mod3}). That is, the treatment effect is constant across $X_j$, but the items constituting the outcome make the observed treatment effect \textit{appear} heterogeneous across $X_j$. Put simply, IL-HTE creates an identification problem for treatment-by-covariate interaction effects when the analysis focuses on the score-level outcome alone.

To illustrate conceptually, imagine two educational interventions, each impacting a construct like math that consists of several subdomains ranging in difficulty (e.g., starting with basic addition and then progressing to multiplication and division of fractions). In the first intervention, the treatment helps previously low-achieving students the most, and this is true across all items. Such a pattern of treatment effects could be caused by, for example, a remedial intervention that provides more tutoring to low-ability students. In this scenario, there is true HTE by baseline achievement, with initially low-achieving students benefiting the most. In the second, there is no HTE by baseline achievement, only IL-HTE: the treatment improves all students' capacity to correctly answer the \textit{easiest} items the most. This pattern could be caused, for example, by a basic skills curriculum provided to all students that improves accuracy rates on easy items but not on hard ones. These two scenarios have distinct implications for policy and how the intervention should be targeted or refined, and yet it can be impossible to distinguish between them when using summary scores alone. Thus, we caution against using a sum score as an outcome variable in a conventional analysis \citep{flake2017construct, mcneish2020thinking, mcneish2024practical} when HTE is of interest, given that the pattern of results produced by each intervention is empirically indistinguishable \citep{spoto2023empirical}.

\begin{figure}
    \centering
    \includegraphics[width=1\linewidth]{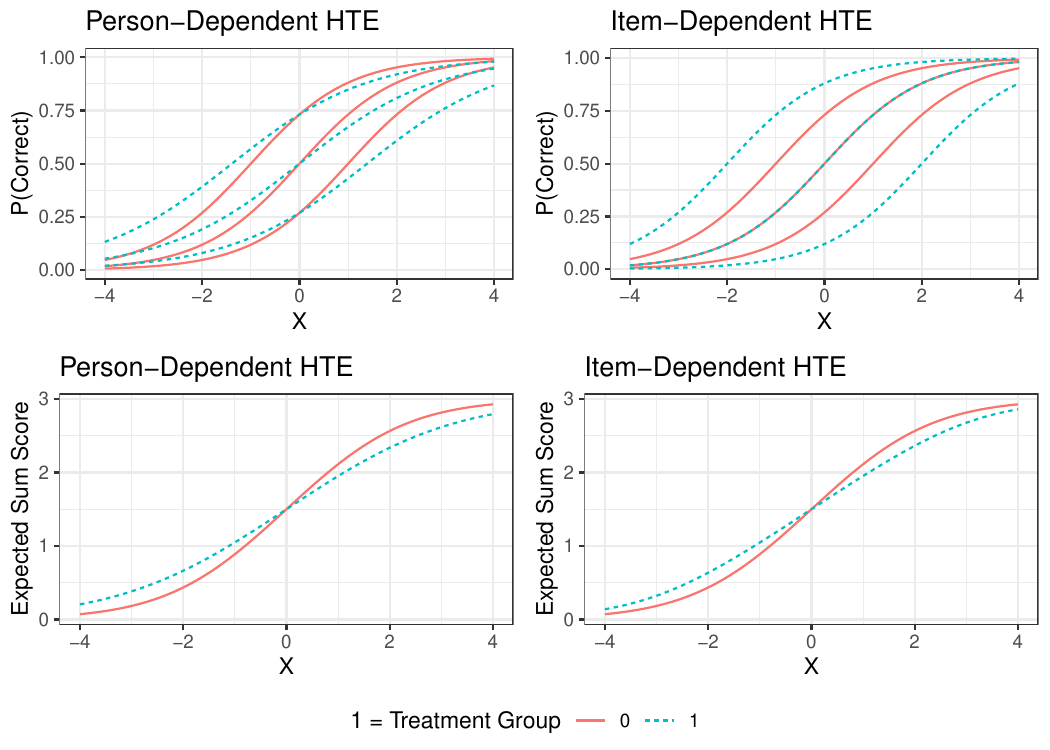}
    \caption{Toy example showing how person-dependent and item-dependent HTE become confounded when the outcome is a sum score}
            \justify \footnotesize
Notes: The top row presents probabilities of a correct response for each item and the bottom row sums the item curves to generate the expected sum score for each test. The horizontal axis represents some pretreatment covariate $X_j$ correlated with the outcome (\textit{not} post intervention ability, or $\theta_j$, as is typical in these types of plots). In the upper left panel, the treatment changes the slopes of the item curves with respect to the covariate $X_j$ and summing these curves to create the test score yields a similar pattern. In the top right panel, the treatment effect is correlated with the item location, yielding a nearly identical pattern in the bottom right panel. (Figure adapted from \cite[p. 82]{gilbert2024disentangling}.)
    \label{fig:trfs}
\end{figure}

However, we can leverage item-level outcome data with the IL-HTE model to solve this identification problem and identify the relevant data-generating process \citep{gilbert2024disentangling}. We illustrate this point conceptually, visually, and empirically with minor mathematical detail; we refer readers interested in a full treatment of this issue to prior publications that include a proof of a simple case and a Monte Carlo simulation study (ibid.). A toy example of the identification problem using a three-item test is displayed in Figure \ref{fig:trfs} \citep{gilbert2024disentangling}. The upper row shows the probability of a correct response to each item for each group as a function of the baseline covariate $X_j$. The left panel shows ``person-dependent'' HTE, namely, an interaction between treatment status and $X_j$, as shown by the different slopes of the item-specific curves by group. When we sum these curves to calculate the expected test score in the bottom left panel, we see a pattern in the overall test score that is the typical signature of an interaction effect. In contrast, the top right panel shows ``item-dependent'' HTE, where $\rho=1$. That is, the easiest item (i.e., the leftmost item) shows a large positive effect, the middle item shows no treatment effect, and the hardest item shows a large negative effect. In essence, the combination of $\sigma_\zeta > 0, \rho=1$ has stretched the item-specific curves with respect to $X_j$ by increasing the SD of the item locations, $\sigma_b$. When we sum these curves in the bottom right panel, the result is an essentially identical pattern in the sum score to the pattern in the bottom left panel. In other words, despite the different underlying data-generating processes, the sum scores displayed in the bottom row are empirically indistinguishable, which presents a serious problem for the interpretation of interaction effects on outcomes derived from psychometric instruments. However, the item-level patterns are quite distinct, suggesting that they can be identified with an appropriate analysis of item-level data. Previous simulation studies have confirmed that these two processes can become confounded, but a flexible model that allows for both person- and item-HTE (Equation \ref{eq:flexible}) eliminates bias and identifies the correct DGP \citep{gilbert2024disentangling, gilbert2024ssri}.\footnote{Tests in the online supplement of \cite{gilbert2024disentangling} show that a linear probability model or factor analytic approach that allows for IL-HTE fails to resolve the identification problem identified here. Furthermore, linear models generally perform poorly when estimating interaction effects for dichotomous or ordinal outcomes \citep{domingue2022ubiquitous}.} We present a more formal mathematical expression of this issue in Appendix \ref{interaction_derivation} and a detailed example in a single empirical dataset in Appendix \ref{interaction_kim}.

\begin{figure}
    \centering
    \includegraphics[width=1\linewidth]{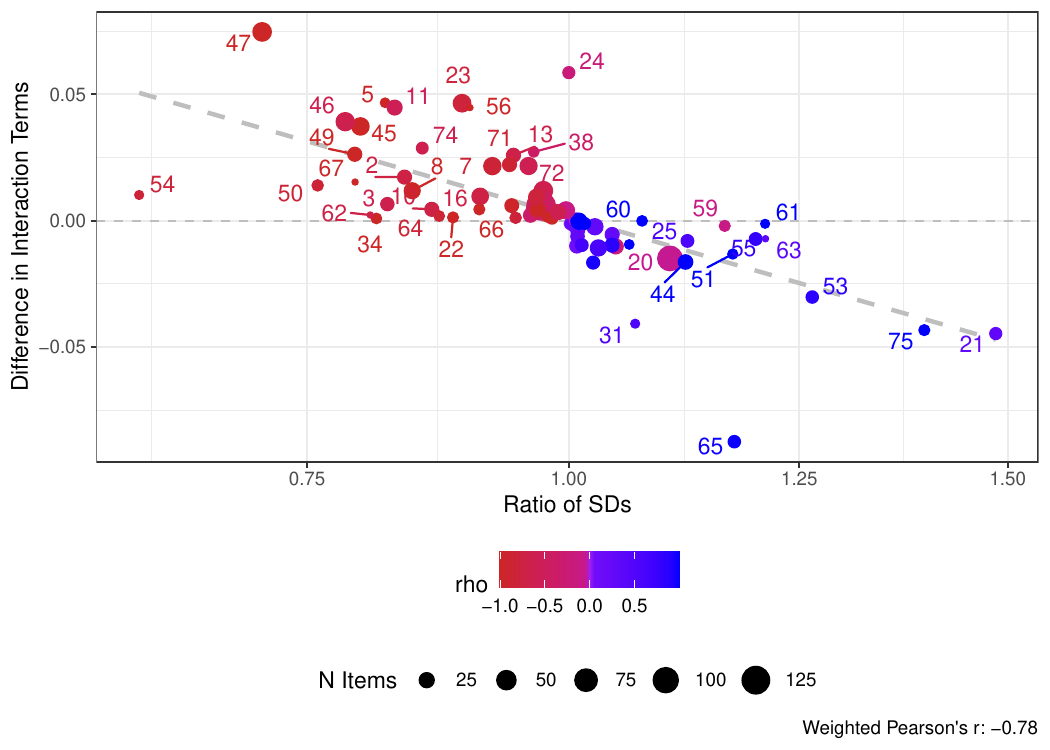}
    \caption{Difference in estimated treatment by baseline interaction terms by dataset}
        \justify \footnotesize
Notes: The vertical axis shows the difference in interaction terms from a model that assumes constant item effects ($\beta_3$ in Equation \ref{eq:mod4}) compared to a model that allows for IL-HTE ($\beta_3$ in Equation \ref{eq:mod5}). The horizontal axis shows the ratio of standard deviations of item locations in the treatment group ($\sigma^*_b$) to the standard deviation in the control group ($\sigma_b$). The points are labeled by dataset ID and color coded by the correlation between item easiness and treatment effect size. Tables of the full model output for each dataset are included in our supplement.
    \label{fig:rho_empirical}
\end{figure}

How large a problem is the potential confounding of interaction effects through treatment-item easiness correlations empirically? \textcite{gilbert2024disentangling} left this as an open question, given the relative paucity of item-level causal analyses examining this question, though at least one study suggests $\rho\neq0$ in an analysis of item-level data from 15 RCTs of educational interventions \citep[Table 1]{ahmed2023heterogeneity}. We explore the phenomenon in our datasets in Figure \ref{fig:rho_empirical}, which plots the change in the estimated treatment-by-baseline interaction term from a model that does not allow for $\rho$ (Equation \ref{eq:mod4}) against one that does (Equation \ref{eq:mod5}) against the ratio of the SD of item locations in the treatment group (denoted $\sigma_b^*$) to the SD of item locations in the control group ($\sigma_b$) in each dataset. According to the mathematical arguments presented in Appendix \ref{interaction_derivation}, a ratio of 1 (i.e., no IL-HTE) will yield no difference in interaction terms, values less than 1 will yield a positive difference, and values greater than 1 will yield a negative difference.

Figure \ref{fig:rho_empirical} provides three key insights. First, the empirical data are in excellent alignment with the theoretical arguments presented above and prior simulation studies showing that IL-HTE creates bias in treatment-by-covariate interaction effects. That is, as the ratio of SDs deviates from 1, the larger the difference between the interaction terms estimated from each model, in a direction that aligns with our theoretical arguments ($r = -.78$). Second, $\rho$ varies widely in empirical data, ranging from near perfect negative correlations to near perfect positive correlations in our data. Third, $\rho$ appears to be a relatively stable feature of interventions. For example, the three outcomes from \cite{duflo2024experimental}, show $\rho > 0$, suggesting that the treatment effects were largest on the easiest items, and this was true across all outcomes. In contrast, studies by Kim et al., who examined variations of the Model of Reading Engagement (MORE) intervention, show $\rho<0$, suggesting that MORE helps more difficult content to a greater extent than easier content. In sum, given an observed treatment-by-baseline covariate interaction effect on a total score (e.g., a sum or IRT-based score), researchers should be quite cautious in interpretation when item-level data are not available and consider the possibility of IL-HTE and treatment-by-item easiness correlations as a potential explanation.\footnote{It is also possible that IL-HTE could suppress a true interaction effect if the difference in slopes and the SD of item locations in the treatment group are working in opposite directions.}

As an example of how these issues could affect the design of policies and interventions, consider school accountability systems. Some evidence suggests that school accountability policies can create incentives for teachers to focus on specific content or students to move as many students as possible above a predefined ``proficiency'' threshold that is used as an accountability metric \citep{booher2005below, wong2011games, lang2010measurement, dee2019causes, macartney2018teacher, payne2010no, dee2011impact, dee2011rules}. In this case, understanding whether a treatment improves the overall performance of low-ability students or improves the accuracy on the easiest items that best distinguish among low-ability students is an important distinction that is not easily addressed without item-level data.

\subsection*{The IL-HTE Model Provides Estimates of Standardized Effect Sizes Corrected for Attenuation Due to Measurement Error} \label{unbiased}

Many studies of assessment or survey outcomes report standardized effect sizes because they are more interpretable and comparable across different measures \citep{schielzeth2010simple}. An underappreciated challenge of using standardized outcome variables is that they yield effect sizes that are biased toward zero by measurement error. This occurs because when we divide a regression coefficient by the pooled SD of the outcome variable $Y_j$ to calculate an effect size such as Cohen's $d$, the estimated SD of $Y_j$ is too large due to measurement error. In particular, the standardized effect size derived from an observed outcome score will be biased toward zero by a factor of $\sqrt{\rho_{YY'}}$, where $\rho_{YY'}$ is the reliability of the measure. The bias can therefore be corrected by dividing the standardized effect size by the square root of an estimate of reliability such as Cronbach's $\alpha$ \citep{gilbert2024measurement, shear2024measurement, hedges1981distribution}.\footnote{Crucially, this bias can still occur in more complex IRT or factor analytic scoring procedures when one produces scores then estimates regressions in separate steps \citep{gilbert2024measurement, stoetzer2022causal, briggs2008using, shear2024measurement}. Alternative IRT scoring procedures such as multigroup models can also address attenuation bias; see \textcite{soland2022survey} for a review.} Combining this fact with the inflation of SEs due to IL-HTE suggests the troubling result that models of psychometric outcomes may be, on average, understating magnitudes but overstating precision, thus yielding a more precise view of a biased estimate. In particular, various studies demonstrate that measurement model misspecification (namely, using a measurement model that does not match the experimental or quasi-experimental study design, or in the present case, assuming a constant effects model when IL-HTE is the true data-generating process) combined with the scoring approach used can affect Type I error rates and power due to the score variances that such model-by-scoring approach interactions produce \citep{soland2022survey, soland2023item, soland2023regression}.

\begin{figure}
    \centering
    \includegraphics[width=1\linewidth]{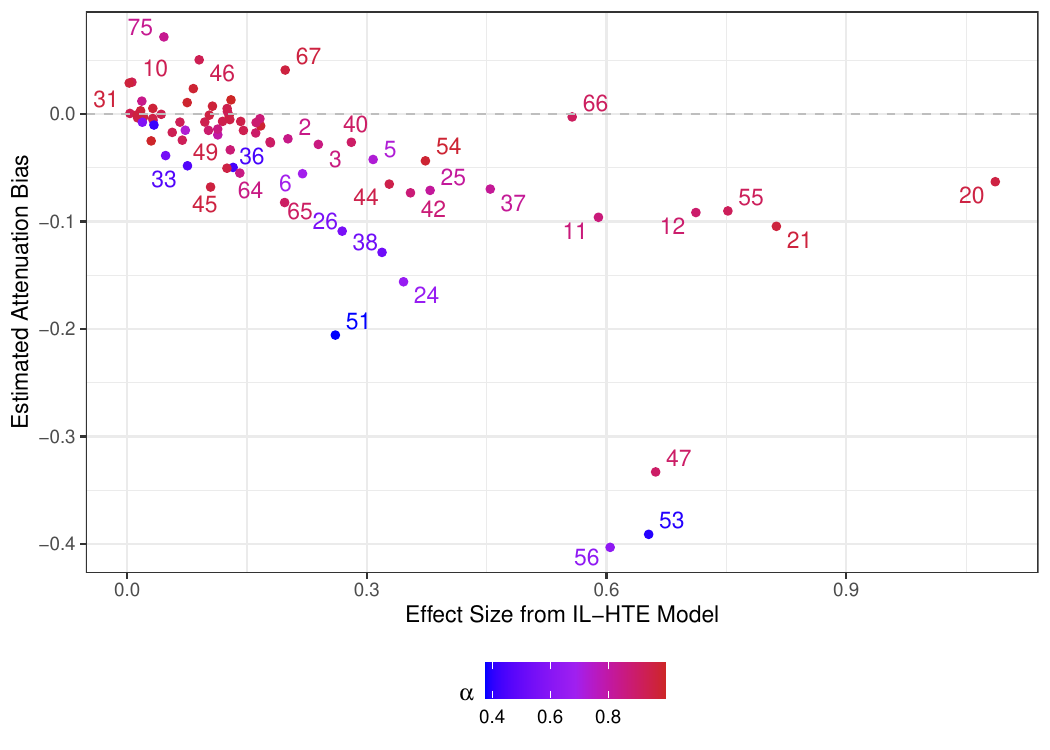}
    \caption{Attenuation of standardized effect sizes due to measurement error}
            \justify \footnotesize
Notes: The figure shows the attenuation bias that arises from measurement error when the outcome variable is standardized. The horizontal axis shows the absolute value of the standardized effect size derived from the IL-HTE model (Equation \ref{eq:mod3}). The vertical axis show the difference between the effect size derived from the IL-HTE model and an effect size estimated by first scoring the outcome with a 1PL IRT model and using this score as the outcome variable in a subsequent regression model. The points are labeled by dataset ID and color coded by the internal consistency of the measure ($\alpha$). We include a table comparing the estimates and standard errors from each model in our supplement.
    \label{fig:attenuation}
\end{figure}

As a latent variable model, the IL-HTE model mitigates this problem because the random effects variances and the fixed effects are simultaneously estimated and appropriately account for the measurement error in the outcome variable. In other words, $\widehat\sigma_\theta$ from the model is a consistent estimator for $\sigma_\theta$ as $N \rightarrow \infty$ (in general and in the presence of IL-HTE), in contrast to $\widehat\sigma_Y$, which is consistent only when both $N \rightarrow \infty$ and $I \rightarrow \infty$. As a result, standardized effect sizes derived from latent variable models are not biased toward zero by measurement error and are therefore more comparable across different tests of varying reliability, or forms of the same test with different numbers of items, which has critical implications for, for example, meta-analyses that combine results from studies using various outcome measures \citep{borenstein2009introduction, gilbert2024mechanisms}. We demonstrate attenuation of the standardized effect sizes due to measurement error in our empirical data in Figure \ref{fig:attenuation}. The vertical axis shows the absolute value of the difference between the effect size derived from Equation \ref{eq:mod3} and the effect size derived from a model with an estimated latent outcome score derived from a 1PL IRT model as the outcome variable, plotted against the effect size derived from Equation \ref{eq:mod3}. We can see the attenuating forces of measurement error in that the effect sizes using the estimated latent outcome scores are consistently smaller in magnitude than those derived from Equation \ref{eq:mod3}, in some cases exceeding .10$\sigma$, a large difference.

\section*{DISCUSSION} \label{discussion}

Traditional methods for HTE analysis that focus on treatment-by-person characteristic interactions are critical, but typically ignore the HTE that may exist among items of an outcome measure. To address this methodological shortcoming, IL-HTE models that allow for unique treatment effects on each assessment item offer an attractive approach. Our evaluation of 5.8 million item responses from 75 datasets from 48 RCTs demonstrates the advantages of the IL-HTE model and the benefits it provides for inference, identification, and generalizability, all essential components of rigorous program evaluation and policy analysis. In short, the IL-HTE model provides an interpretable measure of item-level treatment variation, standard errors that are more theoretically aligned with many applications, estimates of generalizability of treatment effects to untested items, the resolution of critical identification problems in the estimation of interaction effects, and standardized treatment effect sizes that are corrected for attenuation due to measurement error.

We maintained focus on the simple case of individual randomization with two groups at one time point for clarity and to illustrate the affordances of the IL-HTE model, but note that the model can be easily extended to more diverse applications. For example, the IL-HTE model can easily be extended by including additional covariates, randomization blocks, polytomous or continuous items\footnote{When the items are truly continuous, the identification problem for interaction effects is resolved, which we can see by imagining Figure \ref{fig:trfs} with lines instead of curves.} \citep{gilbert2024ssri, bulut2021estimating}, additional levels of hierarchy such as students nested within schools \citep{gilbert2023jebs}, longitudinal structures \citep{gilbert2024ipd}, or fully Bayesian approaches that would allow for more direct claims about the structural parameters of the causal model \citep{burkner2021brms, gilbert2023tutorial, gilbert2024polytomous}.

One constraint of the models explored here is that they assume that each item is equally discriminating with respect to a single latent outcome. That is, they are unidimensional Rasch or one-parameter logistic (1PL) IRT models. The IL-HTE model can be extended to the case of varying discriminations (i.e., a 2PL IRT model) or multidimensionality, although 2PL model implementation requires more advanced software, more computational power, and introduces some interpretational complexities \citep{burkner2021brms, muthen2017mplus, rockwood2019estimating, gilbert2023tutorial, de2014multidimensional, gilbert2024polytomous, gilbert2025network}. Previous simulation studies show that estimation of the ATE in the 1PL IL-HTE model is relatively robust to a 2PL data-generating process \citep[pp. 906-907]{gilbert2023jebs}, and that item discriminations that function as optimal scoring weights are not particularly consequential in causal inference applications where all subjects respond to the same set of items and only the ATE is of interest \citep{gilbert2024measurement}. However, 2PL models are relevant for IL-HTE analysis because, if item discriminations ($a_i$) vary, a treatment that impacts only $\theta_j$ would theoretically manifest as IL-HTE in a misspecified 1PL model because the treatment-control difference on each item would be $a_i \beta_1$ on the logit scale, and thus vary across items even if $\sigma_\zeta = 0$. To examine this issue in our data, we fit 2PL IL-HTE models to our datasets in our supplement and compare them to the 1PL models. We see that $\widehat\sigma_\zeta$ is generally similar across both 1PL and 2PL models ($r = .92$) and that in most cases $\widehat\sigma_\zeta$ is \textit{larger} in the 2PL model, suggesting that the observed IL-HTE in our sample is not driven by model misspecification and that our results may in fact be conservative. We view 2PL extensions of the IL-HTE model as an important area of future methodological research \citep{halpin2024testing}. We include the equation for a 2PL IL-HTE model and some additional discussion in Appendix \ref{subscale} and code to fit such a model in Appendix \ref{r_code}. The impact of multidimensionality on IL-HTE is less studied and similarly provides a promising area of future research that may complement the approach to subscale effects explored here \citep{liao2024curvilinearity, briggs2012value}. 

In short, while more complex models are available, the marginal gains to the accuracy of the results may be limited in many empirical applications \citep{domingue2022intermodel, castellano2015practical, gilbert2024measurement, sijtsma2024sum}. Therefore, given its novelty, the 1PL IL-HTE model presented here has much to offer fields where item response data is commonly collected but rarely used. Furthermore, beyond RCTs, extensions of the IL-HTE model are applicable to alternative experimental and quasi-experimental designs, such as longitudinal growth models, regression discontinuity (RD), multisite trials, and difference-in-differences (DID) designs, underscoring its potential utility in many empirical contexts \citep{soland2023regression, soland2023item, gilbert2024ipd, kuhfeld2023scoring}. We include sample code to fit RD and DID specifications of the IL-HTE model in Appendix \ref{r_code}.

We acknowledge three primary limitations of our approach. First, item-level outcome data is not always available in secondary analysis. For example, in our search for datasets to include in our study, we found several examples of large-scale RCTs that would otherwise have met our inclusion criteria, except that they provided only summary score outcomes. As such, one implication of our results is that researchers should share item-level data in their replication packages \citep{domingue2023item}. Second, poorly designed assessment instruments could yield results that could be mistaken for IL-HTE. For example, if a single item is too easy (or too hard) such that every student in the treatment and control group gets the item right (or wrong), the observed treatment effect on that item would be 0, a result that could be mistaken for IL-HTE even if the true data-generating process is a constant treatment effect model. More generally, poor instrument design can create challenges for all causal identification, not just IL-HTE, for example, when floor or ceiling effects on items mask changes in the underlying latent outcome \citep{domingue2022ubiquitous}. Finally, the IL-HTE model requires large sample sizes and as such is best suited for relatively large data-analytic contexts, with previous simulation studies suggesting minimum sample sizes of approximately 500 subjects and 20 items \citep{gilbert2023jebs, gilbert2024ipd}.

As a final note, the IL-HTE model and latent variable models are generally not nearly as widely used in causal, econometric, or policy evaluation applications as they are in psychology, education, and psychometrics, and as such the explanation and justification of such models may be a difficult task depending on the audience. However, we argue that such a task is a worthy one, given the affordances of latent variable models for causal analyses explored here. Furthermore, evidence from simulation studies shows that the explanatory item response model is in general relatively robust to model misspecification such as skewness or heteroskedasticity in the latent outcome and missing item response data when estimating the ATE \citep{gilbert2023estimating}, and our supplemental analyses show that the IL-HTE model provides similar results regardless of the use of 1PL or 2PL parameterizations with our empirical data. Thus, researchers can be confident in applying the IL-HTE model across a wide range of empirical circumstances. Communicability can be aided by the techniques described in this study, for example, by standardizing coefficients to create findings analogous to standardized effect sizes of sum score outcomes, with the added benefit of correcting for attenuation due to measurement error.

In conclusion, item-level outcome data represent a vast and mostly untapped resource for applied causal inference and the estimation of treatment effect heterogeneity. While item-level outcome data are ubiquitous in education, psychology, epidemiology, and economics, collapsing the item responses to a single summary score obscures important insights into the nature of treatment effects. Using the IL-HTE model, analysts in all quantitative disciplines can gain more valuable insight into for whom and on which items treatments are effective.

\printbibliography

\clearpage

\appendix
\section*{APPENDICES}
\renewcommand{\thesubsection}{\Alph{subsection}}

\subsection{Extending the IL-HTE Model to Allow for Subscale Effects or Varying Item Discriminations} \label{subscale}

\setcounter{table}{0}
\setcounter{figure}{0}
\renewcommand{\thetable}{A.\arabic{table}}
\renewcommand{\thefigure}{A.\arabic{figure}}

\subsubsection{Subscale Effects}

Here, we allow for systematic HTE across subscale $S$, for example, algebra vs. geometry subscales on a math test. For conceptual clarity, we imagine the indicator variable representing subscale, $S_i$, to be mean-centered in a test with an equal number of items in each subscale (e.g., contrast coded as -0.5 for algebra, 0.5 for geometry). In this model, $\beta_1$ continues to reflect the ATE on $\theta_j$, as in  previous models. $\gamma_1$ is a main effect for differences in item easiness between subscales, and $\delta_1$ provides the difference in CATE between the subscales \citep[p. 4]{gilbert2024ssri}. For example, a model in which $\beta_1 = .5, \delta_1 = .5$ would suggest that the treatment improves math proficiency by .5 logits overall, but this effect is the average of an effect of .25 for algebra items and .75 for geometry items. Accordingly, in this model, $\sigma^2_\zeta$ represents \textit{residual} IL-HTE not explained by systematic HTE by subscale, and $b_i^*$ and $\zeta_i^*$ represent the residual item easiness and item-specific treatment effect, respectively. $\rho$ represents the correlation between item easiness and item-specific treatment effect size conditional on subscale. (If $S_i$ is dummy coded instead of contrast coded, $\beta_1$ would represent the CATE on the reference subscale and $\beta_1 + \delta_1$ would represent the CATE on the focal subscale.) For additional empirical examples of IL-HTE subscale analysis, see our references \citep{gilbert2023jebs, gilbert2024ipd, gilbert2024ssri, kim2023longitudinal, gilbert2023tutorial}.

\begin{align}
\label{eq:subscale_mod}
    \text{logit}(\Pr(Y_{ij} = 1)) = \eta_{ij} &= \theta_j + b_{i} + \zeta_i T_j \\ 
    \theta_j &= \beta_0 + \beta_1 T_j + \varepsilon_j \\
    b_i &= \gamma_1 S_i + b_i^* \\
    \zeta_i &= \delta_1 S_i + \zeta_i^* \\
    \begin{bmatrix}
        b_i^* \\
        \zeta_{i}^*
    \end{bmatrix}
     &\sim N\left(\begin{bmatrix}
         0 \\ 0
     \end{bmatrix},\begin{bmatrix}
         \sigma^2_b & \rho\sigma_b\sigma_\zeta \\
         \rho\sigma_b\sigma_\zeta & \sigma^2_\zeta
     \end{bmatrix}\right) \\
    \varepsilon_j &\sim N(0, \sigma^2_\theta).
\end{align}

\subsubsection{2PL Model}

Here, we allow for varying item discriminations by including the discrimination parameter $a_i$ in the model, which is assumed to be equal to 1 in the 1PL formulation. We use a random effects specification for $a_i$, in which $\gamma_0$ represents the discrimination of the average item and $\nu_i$ is the deviation from the average for item $i$, to match our random effects specification for the $b_i$. We use a log link function for the $a_i$ because this constrains the $a_i$ to be positive, imposes a log-normal distribution on the $a_i$, and provides benefits for the stability of model estimation. For identification purposes, $\sigma^2_\theta$ is generally fixed to 1. Alternatively, we could fix $\gamma_0$ to 0 (so that the average $a_i$ is $e^0 = 1$) so that the results are on the same scale as the 1PL model. We use the latter approach in our supplemental analyses for this reason. Under this parameterization, $\beta_1 e^{\gamma_0}$ is the treatment-control difference on the average item, and $(\beta_1 + \zeta_i)(e^{\gamma_0 + \nu_i})$ is the treatment-control difference for item $i$, both on the logit scale. The item random effects may be correlated, such that $\rho_1, \rho_2$ and $\rho_3$ represent the correlations between item easiness and item-specific treatment effect, item easiness and item discrimination, and item-specific treatment effect and item discrimination, respectively.

The 2PL formulation of the IL-HTE model raises some additional interpretational complexities, as even when $\sigma_\zeta = 0$, treatment-control differences will vary across items on the logit scale due to the varying $a_i$. In our view, so long as the ATE $\beta_1$ is fully mediated by $\theta_j$, we do not consider varying item-specific differences that mechanically arise from varying $a_i$ to represent IL-HTE. Rather, we view the residual item-specific effects after accounting for $\beta_1a_i$ to represent IL-HTE. Thus, in the 2PL model, $\sigma_\zeta$ provides an estimate of the SD of the item-specific effects on the linear predictor $\eta_{ij}$ after the expected differences mechanically arising from varying $\beta_1a_i$ have been accounted for. We can also consider this issue from a directed acyclic graph (DAG) perspective. The logic of our DAG (Appendix \ref{dag}) is unchanged whether the paths from the linear predictor $\eta_{ij}$ are fixed at 1, as in the 1PL model, or allowed to vary in a 2PL model.  We thank the second anonymous reviewer for bringing these subtle points to our attention.

See our references for additional resources on this modeling approach and Appendix \ref{r_code} for code to fit a version of this model in R using \texttt{brms} \citep{gilbert2024polytomous, petscher2020past, cho2014additive, burkner2021brms, gilbert2025network}.

\begin{align}
\label{eq:2pl_eirm}
    \text{logit}(\Pr(Y_{ij} = 1)) = a_i(\eta_{ij}) &= a_i(\theta_j + b_{i} + \zeta_i T_j) \\
    \theta_j &= \beta_0 + \beta_1 T_j + \beta_2 X_j + \varepsilon_j \\ 
    \text{ln}(a_i) &= \gamma_0 + \nu_{i} \\
    \begin{bmatrix}
        b_i \\
        \zeta_{i} \\
        \nu_i
    \end{bmatrix}
     &\sim N\left(\begin{bmatrix}
         0 \\ 0 \\ 0
     \end{bmatrix},\begin{bmatrix}
         \sigma^2_b &  & \\
         \rho_1\sigma_b\sigma_\zeta & \sigma^2_\zeta \\
         \rho_2\sigma_b\sigma_a & \rho_3\sigma_\zeta\sigma_a & \sigma^2_a
     \end{bmatrix}\right) \\
    \varepsilon_j &\sim N(0, \sigma^2_\theta).
\end{align}

\clearpage

\subsection{Sample R Code to Fit the IL-HTE Model} \label{r_code}

\setcounter{table}{0}
\setcounter{figure}{0}
\renewcommand{\thetable}{B.\arabic{table}}
\renewcommand{\thefigure}{B.\arabic{figure}}

The R code below illustrates how to fit various EIRMs to a dataset with 0/1 outcome \texttt{score}, 0/1 treatment indicator \texttt{treat}, baseline covariate \texttt{std\_baseline}, person identifier \texttt{s\_id}, mean-centered subscale identifier \texttt{subscale}, and item identifier \texttt{item} using the \texttt{glmer} function (generalized linear mixed effects models in R). For clarity, we omit \texttt{data = dataset, family = binomial} from each \texttt{glmer} function call. For further resources, see the replication materials in our supplement (including code appropriate for clustered or blocked designs) or the various EIRM R tutorials listed in the references \citep{gilbert2023tutorial, de2011estimation}. 

For the 2PL IL-HTE model, we use the Bayesian multilevel modeling software \texttt{brms} \citep{burkner2021brms, gilbert2024polytomous}. Note that for the \texttt{brms} priors, we fix the average item discrimination to 1 so results are directly comparable with the 1PL models, which assume a constant discrimination of 1 for all items. Alternatively, we could set the residual SD of $\theta_j$ to 1, which is the more conventional approach in 2PL models. The recently developed R package \texttt{galamm} \citep{sorensen2024multilevel} can also fit 2PL EIRMs, but does not at present allow for random effects for item discriminations that better align with the IL-HTE framework (Ø. Sørensen, personal communication, August 14, 2024). We can however fit a fixed intercepts, random coefficients (FIRC, see \cite{bloom2017using}) parameterization of the 2PL IL-HTE model using \texttt{galamm}; we include the relevant code as a reference but do not pursue this modeling extension further. Note that the FIRC approach does not allow for item characteristics to be included in the model as they are collinear with the item fixed effects.

Extending the IL-HTE model to regression discontinuity and difference-in-differences approaches is straightforward. We present code to fit such models in a 1PL framework below.

\singlespacing
\begin{verbatim}
# load libraries
library(lme4)
library(brms)
library(galamm)

# 1PL IL-HTE model with lme4

# treatment indicator only
glmer(score ~ treat + (1|s_id) + (1|item))

# constant effects model with pretest
glmer(score ~ treat + std_baseline + (1|s_id) + (1|item))

# IL-HTE model
glmer(score ~ treat + std_baseline + (1|s_id) + (treat|item))

# person interaction model
glmer(score ~ treat*std_baseline + (1|s_id) + (1|item))

# flexible model
glmer(score ~ treat*std_baseline + (1|s_id) + (treat|item))

# subscale model
glmer(score ~ treat*subscale + std_baseline + (1|s_id) + (treat|item))

# 2PL IL-HTE model with brms

# set priors
# note that the normal priors on the random effects 
# represent half-normal distributions
prior <- 
  # sd of item easiness
  prior("normal(0,1)", class = "sd", group = "item", nlpar = "eta") +
  # sd of person ability
  prior("normal(0, 1)", class = "sd", group = "s_id", nlpar = "eta") +
  # sd of item discrimination
  prior("normal(0, .5)", class = "sd", group = "item", nlpar = "logalpha") +
  # TE on theta
  prior("normal(0, .5)", class = "b", coef = "treat", nlpar = "eta") +
  # IL-HTE
  prior("normal(0, .5)", class = "sd", coef = "treat", group = "item", nlpar = "eta") +
  # theta intercept
  prior("normal(0, 1)", class = "b", coef = "Intercept", nlpar = "eta") +
  # disc intercept - constant at exp(0) = 1
  prior("constant(0)", class = "b", coef = "Intercept", nlpar = "logalpha") +
  # baseline
  prior("normal(.5, 1)", class = "b", coef = "std_baseline", nlpar = "eta")

# declare the model
mod <- bf(
  # logalpha is the log discrimination parameter
  score ~ exp(logalpha)*eta,
  # model for the linear predictor
  eta ~ 1 + treat + std_baseline + (1|s_id) + (treat|item),
  # model for logalpha
  # item locations and discriminations can be modeled as
  # correlated by using (1|i|item) instead of (1|item)
  logalpha ~ 1 + (1|item),
  # declare non-linear model
  nl = TRUE
)

# fit the model
fit_2pl <- brm(
  formula = mod,
  data = {dataset},
  family = brmsfamily("bernoulli", "logit"),
  prior = prior,
  backend = "cmdstanr",
  chains = 4,
  iter = 2000,
  cores = 4,
  threads = threading(4),
  refresh = 200
)

# 2PL IL-HTE model with galamm

# declare the matrix of item discriminations
# the first discrimination is fixed to 1 for identification
# NA means the discrimination is freely estimated
I <- {number of items here}
mat <- matrix(c(1, rep(NA, I-1)), ncol = 1)

# fit the model
# here we have fixed effects for item intercepts
# fixed item discriminations
# and a random slope for treatment
# "ability" is the arbitrary name of the latent variable
fit_2pl <- galamm(
  formula = score ~ 0 + treat + std_baseline + item + 
            (0 + treat|item) + (0 + ability|s_id),
  data = {dataset},
  family = binomial,
  factor = "ability",
  load.var = "item",
  lambda = mat
)

# Regression Discontinuity
# here, running_var is the running variable centered at the cutoff
glmer(score ~ treat*running_var + (treat|item) + (1|s_id))

# Difference-in-Differences
# here, we assume repeated administration of the items
# at two periods indexed by the indicator variable post
# treatXpost is the interaction between treatment and post period
glmer(score ~ treat + post + treatXpost + (treatXpost|item) + (1|s_id))

\end{verbatim}

\clearpage

\subsection{Algebra for Sensitivity Analysis} \label{sens_algebra}

\setcounter{table}{0}
\setcounter{figure}{0}
\renewcommand{\thetable}{C.\arabic{table}}
\renewcommand{\thefigure}{C.\arabic{figure}}
\doublespacing

Assuming a positive $\beta_1$, we can solve for the point at which the treatment effect becomes statistically insignificant (RI = random intercepts).
\begin{align}
    \widehat{\beta_1} - 1.96 \sqrt{\widehat{V}(\beta_1)_{\text{RI}} + \frac{\sigma^2_\zeta}{I}} &= 0 \\
    - 1.96 \sqrt{\widehat{V}(\beta_1)_{\text{RI}} + \frac{\sigma^2_\zeta}{I}} &= -\widehat\beta_1 \\
    \sqrt{\widehat{V}(\beta_1)_{\text{RI}} + \frac{\sigma^2_\zeta}{I}} &= \frac{\widehat\beta_1}{1.96} \\
    \widehat{V}(\beta_1)_{\text{RI}} + \frac{\sigma^2_\zeta}{I} &= (\frac{\widehat\beta_1}{1.96})^2 \\
     \frac{\sigma^2_\zeta}{I} &= (\frac{\widehat\beta_1}{1.96})^2 - \widehat{V}(\beta_1)_{\text{RI}} \\
     {\sigma^2_\zeta} &= I((\frac{\widehat\beta_1}{1.96})^2 - \widehat{V}(\beta_1)_{\text{RI}}) \\
     \sigma_\zeta &= \sqrt{I((\frac{\widehat\beta_1}{1.96})^2 - \widehat{V}(\beta_1)_{\text{RI}})}
\end{align}

\clearpage

\subsection{Additional Mathematical Detail for the Interaction Effect Identification Problem} \label{interaction_derivation}

\setcounter{table}{0}
\setcounter{figure}{0}
\renewcommand{\thetable}{D.\arabic{table}}
\renewcommand{\thefigure}{D.\arabic{figure}}

We can formalize the intuition provided by Figure \ref{fig:trfs} by noting that the flattening of the sum score curves (i.e., the test response functions, or TRFs) with respect to $X_j$ can result from two distinct processes. First, through a direct flattening of the slopes of the individual item curves, as in the interaction case, and second, through an increase in the variance of the item locations in the treatment group that arises from the combination of $\sigma_\zeta$ and $\rho$. That is, $\sigma_b$ provides the SD of item locations in the control group, and the presence of IL-HTE changes the SD of item locations in the treatment group according to the following formula \citep[pp. 29-32]{steele2008module}, where we define $\sigma_b^*$ as the SD of item locations in the treatment group:

\begin{align}
    \sigma_{b}^* &= \sqrt{\sigma^2_b + \sigma^2_\zeta + 2 \rho \sigma_b \sigma_\zeta}. 
\end{align}
When $\sigma_\zeta=0$, $\sigma_b = \sigma_b^*$ and the identification problem is resolved. Furthermore, because $\sigma_b$ is typically larger than $\sigma_\zeta$, and $\sigma_\zeta$ is typically less than 1 (relative to $\sigma_\theta$, as is the case in our data, as shown in Figure \ref{fig:se_inflation}), when $\rho = 0$, the final term drops out and the resulting inflation of $\sigma^*_b$ is much less severe. For example, consider the simple case where the SD of the item locations in the control group ($\sigma_b$), the SD of item-specific treatment effects ($\sigma_\zeta$), and the location-treatment effect correlation ($\rho$) are all 1, so that $\sigma_b^*$ will be twice as large as $\sigma_b$ (i.e., the treatment doubles the SD of the item locations):

\begin{align}
    \sigma_{b}^* &= \sqrt{\sigma^2_b + \sigma^2_\zeta + 2 \rho \sigma_b \sigma_\zeta}\\
    &= \sqrt{1 + 1 + 2} \\
    &= \sqrt{4} \\
    &= 2.
\end{align}
This increase in $\sigma_b^*$ relative to $\sigma_b$ will attenuate the slope of the sum score curve. We can demonstrate this with the following formula that allows us to convert the slopes of the item curves to the slope of the sum score curve \citep[p. 116]{huang2022analyzing}:

\begin{align}
    \beta_{S} \approx \frac{\beta_{I}}{\sqrt{.346 \sigma^2_b + 1}}, \label{eq:atten}
\end{align}

\noindent where $\beta_S$ is the slope of the sum score curve (on the logit scale) and $\beta_I$ is the slope of the item curve. 

To illustrate, we apply Equation \ref{eq:atten} to both the treatment and control groups, continuing with the simple case where $\sigma_b = 1$ and $\sigma_b^* = 2$ as described above. In the control group, if we assume $\beta_I = 1$, then $\beta_S$ is

\begin{align}
    \beta_{S} \text{ if } (T_j = 0) &\approx \frac{1}{\sqrt{.346 \times 1 + 1}} \\
    &\approx .86,
\end{align}
\noindent and in the treatment group,

\begin{align}
    \beta_{S} \text{ if } (T_j = 1) &\approx \frac{1}{\sqrt{.346 \sigma^{2*}_b + 1}} \\
    &\approx \frac{1}{\sqrt{.346 \times 4 + 1}} \\
    &\approx .65.
\end{align}
\noindent Thus, even though the slopes of the item curves in the numerator are identical, the difference in the slopes of the sum scores is approximately $.65 - .86 = -.21$, which could easily be misinterpreted as an interaction effect between treatment and the baseline covariate in an analysis of the sum scores. 

Similarly, $\rho<0$ will produce a spurious interaction in the other direction by ``shrinking'' the scale inward. Assuming $\rho = -1$,

\begin{align}
    \beta_{S} \text{ if } (T_j = 1) &\approx \frac{1}{\sqrt{.346 \sigma^{2*}_b + 1}} \\
    &\approx \frac{1}{\sqrt{.346 \times (1 + 1 - 2) + 1}} \\
    &\approx 1.
\end{align}

\noindent Here, the difference in slopes of the sum scores is approximately $1 - .86 = +.14$. Thus, for the person-dependent and item-dependent processes to produce approximately the same pattern of sum scores, the following must hold:

\begin{align}
    \frac{\beta_2 + \beta_3}{\sqrt{.346 \sigma_b^2 + 1}} \approx \frac{\beta_2}{\sqrt{.346 \times (\sigma_b^2 + \sigma_\zeta^2 + \rho \sigma_b \sigma_\zeta) + 1}}.
\end{align}

\noindent In the control group, these quantities will clearly be the same as both $\beta_3$ and $\sigma_\zeta$ are 0. For a given set of parameters $\beta_2, \beta_3, \sigma_b$ we can choose infinitely many values of $\sigma_\zeta$ and $\rho$ that produce the equivalent pattern in the total score (to a point; $\rho = -1$ can only compress the scale such that the individual item slope is not attenuated, when $\sigma^{2*}_b = 0$). Clearly, therefore, an observed treatment by covariate interaction in a model of the sum score could result from either a true interaction or a correlation between item-specific treatment effects and item location. Simulations in \textcite{gilbert2024disentangling} show that this phenomenon is not an artifact of sum scores. The results apply equally to IRT-based scores and EIRMs that simultaneously estimate the measurement and structural models.

We emphasize $\rho$ as the primary cause of the identification problem because, while theoretically a large enough value of $\sigma^2_\zeta$ could generate the same problem when $\rho = 0$, $\sigma^2_\zeta$ is typically less than $\sigma^2_b$ and less than 1, so that $\sigma_\zeta > \sigma^2_\zeta$. Furthermore, $\sigma_\zeta$ must be positive, whereas a negative value of $\rho$ can stretch the scale in either direction. As a final point, we note that any bias in interaction terms will necessarily bias main effects in models that omit the interaction, because the main effects are weighted averages of the effects in each subgroup.

\clearpage

\subsection{Illustration of the Interaction Effect Identification Problem with a Single dataset} \label{interaction_kim}

To show what the confounding of person- and item-dependent HTE looks like in a single empirical dataset, we illustrate with data from \cite{kim2021improving}b, who evaluate the effect of the Model of Reading Engagement (MORE) intervention on the vocabulary knowledge of Grade 1 students. The estimated treatment by baseline test score interaction effect is attenuated by about a third, reducing from $.15\sigma$ in the model that does not allow for IL-HTE compared to $.10\sigma$ in the IL-HTE model, a large reduction. The data are shown in Figure \ref{fig:kim_items}, which plots the mean scores (left panel), the correct response probabilities (middle) and the correct response log-odds (right) as a function of treatment status and baseline scores, and the items are ordered from most to least difficult. We see that the apparent treatment by baseline interaction in the mean scores is partially driven by $\rho$ ($\widehat\rho = -.56$), suggesting that the treatment effect is largest on the most difficult vocabulary words, which, coupled with the large value of $\sigma_\zeta$ in this dataset leads to an inflated interaction term, as we can see in the panels showing the item-level data, which show a much smaller difference in slopes compared to the mean scores. An advantage of the flexible model (Equation \ref{eq:flexible}) is that it allows both sources of HTE to be estimated simultaneously, providing the most accurate view of HTE. In this case, we can see that the MORE intervention helped both the highest achieving students the most, on average across all items, \textit{and} helped the most difficult content the most, on average across all students.

\setcounter{table}{0}
\setcounter{figure}{0}
\renewcommand{\thetable}{E.\arabic{table}}
\renewcommand{\thefigure}{E.\arabic{figure}}

\begin{figure}
    \centering
    \includegraphics[width=1\linewidth]{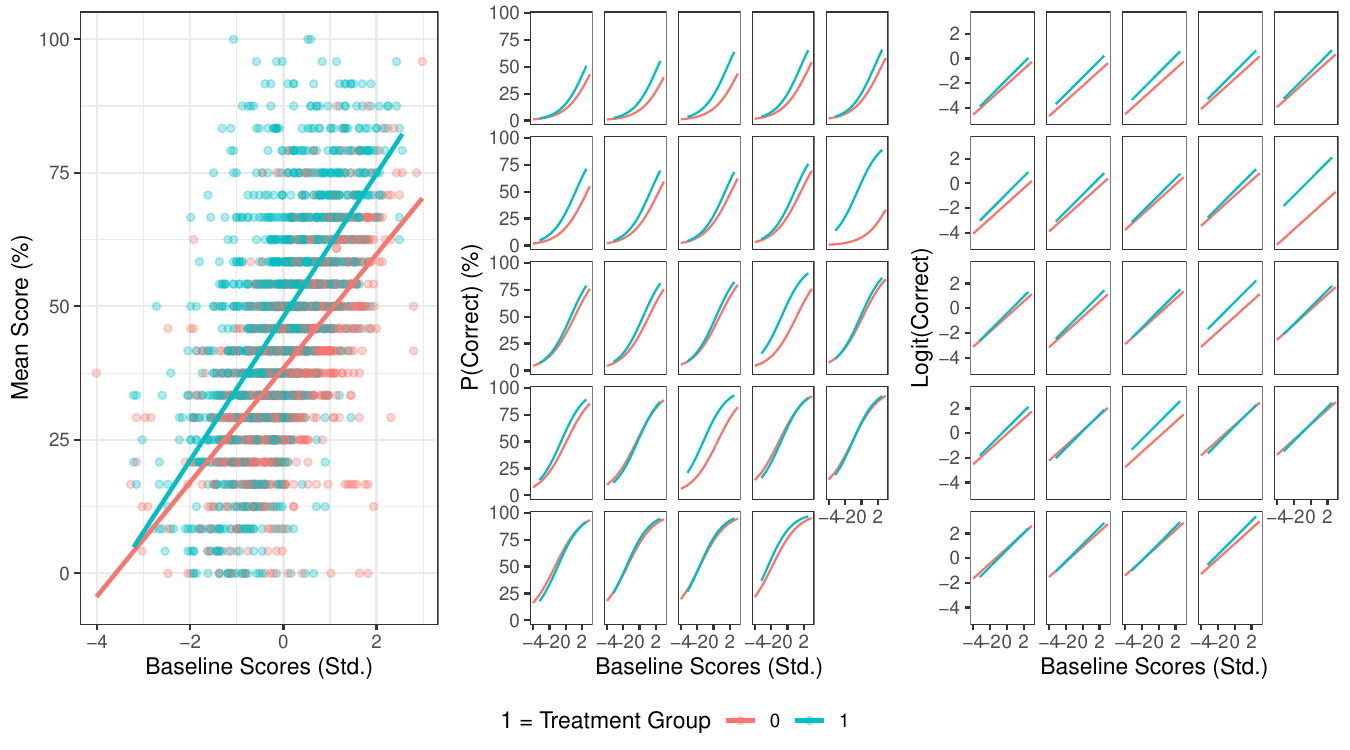}
    \caption{Empirical data from Kim 2021b showing the confounding of person-dependent and item-dependent HTE}
             \justify \footnotesize
Notes: The left panel provides a scatter plot of the mean test score on a researcher-designed vocabulary assessment containing 24 items against standardized baseline scores and shows a large positive interaction between baseline scores and treatment status. The middle panel shows the probabilities of a correct response for each item, ordered from most difficult to least difficult, and we can see that the treatment effects appear to be the largest on the most difficult items. The right panel shows the item responses in log-odds, and we can see the pattern more clearly. While there is a slight difference in the slopes in the log-odds, the large interaction in the mean score is partially driven by the concentration of larger effects on the most difficult items. The item curves are derived from a fixed effects model of the correct response on treatment, item, and baseline score, with two-way interactions between treatment and item and treatment and baseline score. 
    \label{fig:kim_items}
\end{figure}

\clearpage
\subsection{Directed Acyclic Graph of the IL-HTE Model} \label{dag}

\setcounter{table}{0}
\setcounter{figure}{0}
\renewcommand{\thetable}{F.\arabic{table}}
\renewcommand{\thefigure}{F.\arabic{figure}}

\begin{figure}[h]
    \centering
    \includegraphics[width=1\linewidth]{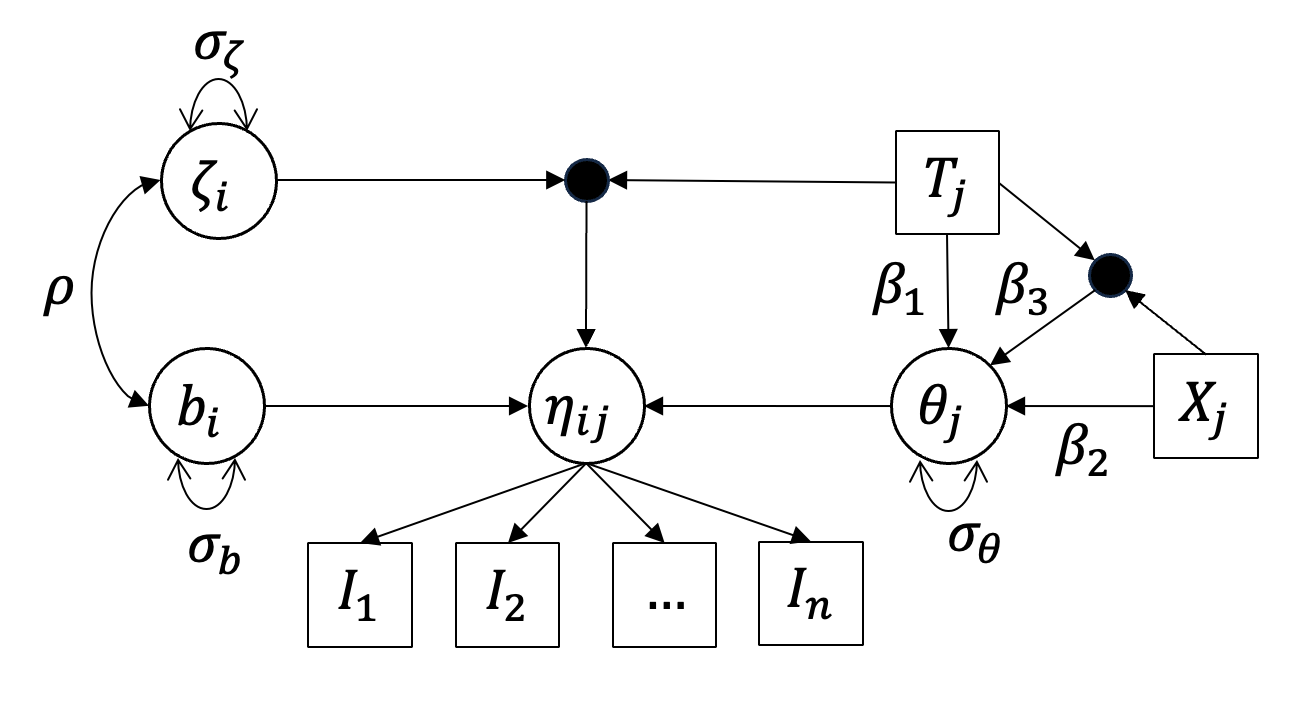}
    \caption{Directed Acyclic Graph for Equation \ref{eq:flexible}}
            \justify \footnotesize
Notes: Squares indicate observed variables, hollow circles indicate latent variables, and solid circles represent cross product interaction terms. $I_n$ are item responses, $T_j$ is the treatment indicator, and $X_j$ is the covariate. $\beta_1$ represents the average treatment effect. $\rho$ represents the correlation between item location and item-specific treatment effect size. Path coefficients are fixed at 1 unless otherwise indicated (a 2PL model would allow the paths from $\eta_{ij}$ to the $I$ to vary). The $\sigma$ terms represent residual standard deviations. We could allow for subscale effects by including item characteristics as predictors of $b_i$ and $\zeta_i$.
    \label{fig:dag}
\end{figure}

\end{document}